\documentclass[conference]{IEEEtran}
\IEEEoverridecommandlockouts
\usepackage{url}
\usepackage[cmex10]{amsmath}
\usepackage{bm}
\usepackage{amssymb,amsthm,mathrsfs,amsfonts} 
\usepackage{mathtools}
\usepackage{verbatim}
\usepackage{color,soul}
\usepackage{graphicx}
\usepackage[linesnumbered,algoruled,boxed,lined]{algorithm2e}
\usepackage{caption}
\usepackage{subcaption}
\usepackage{multicol}

\usepackage{enumitem}
\usepackage{float}
\usepackage{cite}

\newtheorem{defi}{Definition}
\newtheorem{theo}{Theorem}
\newtheorem{lem}{Lemma}

\usepackage{textcomp}
\usepackage{xcolor}
\def\BibTeX{{\rm B\kern-.05em{\sc i\kern-.025em b}\kern-.08em T\kern-.1667em\lower.7ex\hbox{E}\kern-.125emX}}  
  
\usepackage[a4paper,right=0.625in,left=0.625in,top=0.75in,bottom=1in]{geometry}

\def\occ{{\overline{\mathcal{C}}}}

\def\cS{{\mathcal{S}}}

\def\bs{{\bm{s}}}    
\def\cL{{\mathcal{L}}}

\def\l{\ell}
\def\bl{{\bm{\l}}}
\def\sN{[N]}
\def\bI{{\mathbb{I}}}
\def\oP{{\overline{P}}}
\def\omu{{\overline{\mu}}}    
\def\*{{\star}}

\def\10i{\!\!\!\!\!\!\!\!\!\!\!}

\newcommand{\remove}[1]{}
\newcommand\myeq{\stackrel{\mathclap{\normalfont\mbox{def}}}{=}}

\begin{document}

\title{Stability Analysis of Device-to-Device Relay-Assisted Cellular Networks}

\author{\IEEEauthorblockN{Soubhik Deb}
\IEEEauthorblockA{IIT Bombay\\
soubhikdeb1994@gmail.com}
\and
\IEEEauthorblockN{Prasanna Chaporkar}
\IEEEauthorblockA{IIT Bombay \\
chaporkar@ee.iitb.ac.in}
\and
\IEEEauthorblockN{Abhay Karandikar}
\IEEEauthorblockA{IIT Bombay \\
karandi@ee.iitb.ac.in}
}

\maketitle

\begin{abstract}
Motivated by increasing popularity of delay-sensitive applications, we investigate the queue stability in device-to-device (D$2$D) relay-assisted cellular networks. In contrast to prior works on D$2$D relay-assisted cellular networks, we incorporate practical properties of these networks such as bursty packet arrivals, user mobility and relays generating their own traffic. Assuming network topology evolving in IID fashion, we first evaluate the system stability region to quantify its delay performance. Subsequently, we formulate a policy for joint resource allocation and power control under a more realistic mobility scenario of generalized reflected random walk. Also, the throughput optimality of this policy is investigated. Simulation results are provided to give better understanding of queue stability in the network.
\end{abstract}


\section{Introduction} 
Device-to-device (D2D) communications is a promising technique for complementing and enhancing the conventional cellular systems through spatial diversity \cite{JYH}. With D$2$D communications, a user can act as a virtual infrastructure to relay the data being transmitted from other users in the network for combating fading and improving coverage \cite{DLYGSG}. Compared to traditional relays in relay-assisted cellular networks, such D$2$D relay nodes can be deployed without incurring any additional infrastructure cost.

With the rising prospects for deployment of D$2$D relay-assisted cellular networks \cite{PRO}, extensive research is being undertaken on it. In \cite{SN}, the authors formulated analytical models for the uplink coverage probability and spectral efficiency of D$2$D relay-assisted cellular networks. In \cite{JOT}, the authors considered D$2$D relaying in multi-cell downlink networks. They analyzed the inter-cell interference using fluid network model and captured the interaction between relaying decisions and inter-cell interference. In \cite{SXYGW}, the authors investigated joint D$2$D relay node placement and power allocation with the objective of maximizing energy efficiency. In \cite{NVJLM}, the authors studied the problem of incentivizing users to cooperate in relaying. Despite this recent progress, prior works in D$2$D relay-assisted cellular networks have only considered physical layer dynamics. Since most applications are delay-sensitive, it is very important to consider delay performance in addition to physical layer throughput. Addressing resource allocation and power control for delay-sensitive applications is non-trivial as it involves both queueing theory (to model the queue dynamics) and information theory (to model the physical layer dynamics).

The defining feature of D$2$D relay-assisted cellular networks that makes it so attractive is that users act as D$2$D relay nodes. This implicitly infers that relays can generate their own exogenous arrivals to the network which complicates the design of policies for resource allocation and power control. Additionally, mobility of users is a paramount feature in wireless networks which significantly affects the achievable performances \cite{JHNTA}. An example of mobility is the usage of smartphones as relays \cite{HMN}. To the authors' best knowledge, no previous work has investigated the delay performance in the D$2$D relay-assisted cellular networks while including the above practical features.

In this paper, each user is assumed to be mobile with bursty arrivals and has queues of infinite length for storing the arriving packets. Each user acts as both source and relay for other users with the base station (BS) being the final destination for all the packets in the network. The goal of this work is to formulate a policy for resource allocation and power control for an underlay D$2$D relay-assisted cellular network that guarantees queue stability for each user while satisfying average power constraints. For brevity, hereinafter, each user will be referred to as \emph{Mobile Station} (MS). In summary, our contributions are as follows:
\begin{itemize}
\item In contrast to prior works on D$2$D relay-assisted cellular networks, our network includes practical features such as bursty nature of packet arrivals, D$2$D relays generating their own traffic to the network and mobility of all MSs.  
\item We evaluated the \textit{system stability region} to quantify the delay performance under the assumption that the network topology evolves in an IID fashion. 
\item We formulated an \textit{online policy} for joint resource allocation and power control for network topology evolving as a generalized random walk that is proved to be  $\epsilon-$\textit{throughput optimal} in terms of queue stability.
\end{itemize}

The remainder of the paper is organised as follows. We describe the proposed system model in Section~\ref{sec:SysMod}. The system stability region is quantified in Section~\ref{sec:QuantSysRateReg}. In Section~\ref{sec:OnThroPol}, we present the proposed online scheduling policy for joint resource allocation and power control and investigate its optimality in terms of queue stability. In Section~\ref{sec:NumRes}, we present numerical simulation results on the performance of the policy. Finally, we draw the main conclusions in Section~\ref{sec:Con}.

\section{System Model}
\label{sec:SysMod}
\begin{figure}[t]
    \centering
    \includegraphics[width=0.45\textwidth]{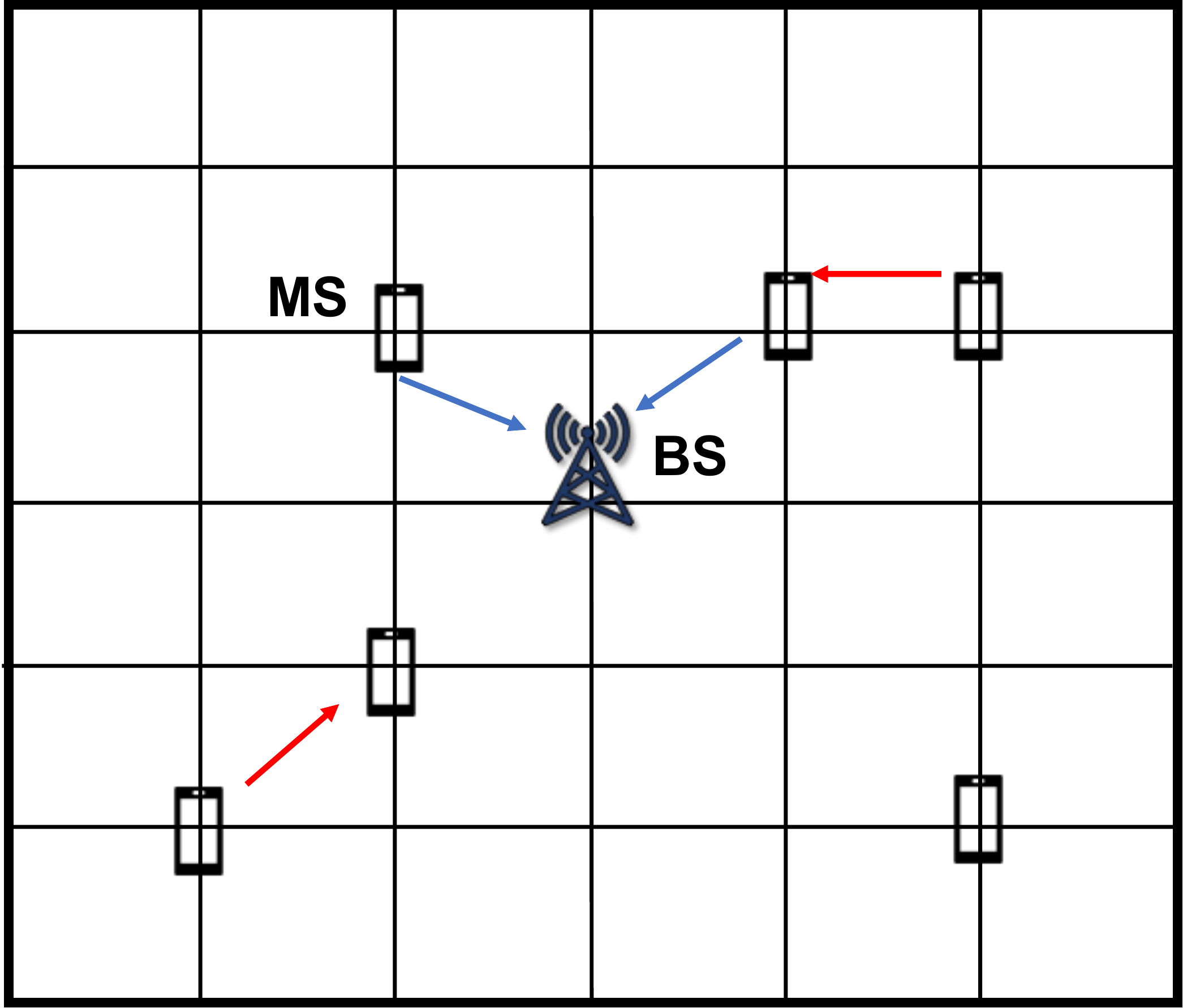}
    \caption{A snapshot of a D$2$D relay-assisted cellular network with BS and MSs in a slot. The blue links are the cellular links and the red links are the D$2$D links. The MSs are equipped with full-duplex tranceivers.}
    \label{fig:Grid}
\end{figure}

We consider a single cell having one BS and $N$ MSs equipped with full-duplex tranceivers. Define $[N] = \{1,\ldots,N\}$ where nodes $1, ..., N$ represent MSs in the network and node $0$ represents the BS. The direct transmission between an MS and the BS is a \textit{cellular link} and that between any two MSs is a \textit{D$2$D link}. In this paper, only uplink is considered and, thus, an MS transmits its data either directly to BS or relay via another MS. For tractability, the number of hops is limited to two. With focus on underlay D$2$D communication, the spectrum is shared between cellular links and D2D links. Time is slotted and each slot has equal length. The coverage area of the BS is divided into a uniform square grid. Each MS $i$ is on a grid point in every slot. Fig.~\ref{fig:Grid} illustrates the system model described here. To model mobility, we assume that each MS performs generalised reflected random walk on the grid points, i.e. each MS stays at the current position or moves to one of the adjacent grid points. The probability with which a MS moves may vary from MS to MS and may also depend on the current position of the MS. However, we assume that the movements of various MSs are independent, and  mobility process for each MS is an irreducible and aperiodic Markov chain. Let $s(n)$ denote the network topology in slot $n$. Let $\cS$ denote the set of all possible network topologies, i.e. $s(n)\in\cS$ with probability $1$ for every $n$. Also, let $[\pi_s]_{s\in\cS}$ denote the steady state distribution of the Markov chain modelling the evolution of network topology.

At each node $i \in [N]$, exogenous packets arrive as per arrival process $\{A_{i}(n)\}_{n=1}^{\infty}$,  where $A_{i}(n)$ is the number of packets arriving at MS $i$ in slot $n$. We assume that the arrivals are independent and identically distributed (IID) across slots for every MS $i$. Moreover, arrivals are independent across MSs. Let $A_{i}(n) \le A_{\max}$ for every MS $i \in [N]$. Also, let $\lambda_{i}=\mathbb{E}[A_{i}(n)]$ and $\bm{\lambda}=[\lambda_{1} \ \cdots \ \lambda_{N}]$ denote the arrival rate vector. At each MS, we allow for transmit power selection and thereby rate adaptation. Specifically, we assume that each MS can pick one of the finitely many possible power levels available to it. A transmit power vector $[\l_{10}, \l_{12}, ...,\l_{ij}, ...]$ is an $N^2$-dimensional vector in which $(i,j)^{\rm th}$ element, $\l_{ij}$, for $i,j \in \sN$, denotes the power expended by MS $i$ to transmit packets to MS $j$ over the link $(i,j)$ and $\l_{i0}$ is the power expended by MS $i$ for transmitting directly to BS. Let $\cL$ denote the set of all possible power vectors satisfying the assumption that in each slot, an MS can transmit to only one receiver. Denote by $\mu_{ij}(s,\bl)$ the transmission rate (in packets per slot) on the link $(i,j)$ in network topology $s \in \cS$ with power vector $\bl \in \cL$. Consider $\mu_{max}$ be the maximum possible transmission rate across any link in the network at any topology. Additionally, suppose that each MS $i$ complies with average power constraint. That is,
\begin{align}
\label{eq:PowCons}
\lim_{T\to\infty}\frac{1}{T}\sum_{n=1}^{T}\sum_{j \in [N]\cup\{0\} \setminus \{i\}}\l_{ij}(n) \leq \oP_i.
\end{align}

Our proposed network entails each MS to maintain two queues, \textit{own queue} for storing packets arriving exogenously to the MS and \textit{relay queue} to store the packets arriving from other MSs for relaying to BS. With number of hops limited to two, packets from the \textit{relay queue} are transmitted directly to the BS. Let $X_i(n)$ and $Y_i(n)$ represent the backlog in the \textit{own queue} and \textit{relay queue} of MS $i$ in slot $n$, respectively. The queues' dynamics are described as follows: For every $i \in \mathbb{N}$ and $n \in \mathbb{Z}^{+}\bigcup\{0\}$,
\begin{align}
\label{eq:X_evol}
    X_{i}(n+1)&=\max\{X_{i}(n)- \sum_{j \in \sN}\mu_{ij}(s(n),\bl(n))  \nonumber \\
    & \hspace{1cm}-\mu_{i0}(s(n),\bl(n)),0\} + A_{i}(n) \\
\label{eq:Y_evol}    
    Y_{i}(n+1)&=\max\{Y_{i}(n)-\mu_{i0}(s(n),\bl(n)),0\} \nonumber \\
    &\hspace{1cm} + \sum_{j\in\sN}\mu_{ji}(s(n),\bl(n)). 
\end{align}

Additionally, on the lines of \cite{NE}, we establish a \textit{virtual power queue }  for each MS $i$ with queue length $U_{i}(n)$ equal to the maximum excess power expended beyond the average power constraint over any interval ending in slot $n$. Defining $U_{i}(0)=0$, its dynamics evolve as follows:
\begin{align}
\label{eq:dynU}
U_{i}(n+1)=\max\{U_{i}(n)-\oP_{i},0\}+\sum_{j=0}^N\l_{ij}(n).
\end{align} 
The $U_{i}(n)$ process acts as a single-server queue with constant departure in each slot given by the average power allocation $\oP_{i}$, with `arrivals' given by the total power allocated for outgoing transmissions of MS $i$ in the current time slot. The intuition behind this construction is that if a scheduling policy $\Delta$ with power vector $\bl^{\Delta} \in \cL$ stabilises queues $X_{i}(n)$, $Y_{i}(n)$ and the virtual queue $U_{i}(n)$ for every $i \in [N]$, then that policy also satisfies the average power constraint for each MS (\ref{eq:PowCons}). Consider $P_{max}$ be the maximum transmit power for any MS in the network.

Due to mobility, an MS can have good channel to the BS in one slot and can has poor channel to the BS in another slot. The MS can conveniently transmit its data to another MS that have good channel with the BS. The second MS can store the received packets in its \textit{relay queue}. The second MS can later transmit the packets buffered in \textit{relay queue} depending on availability of appropriate channel state and queue length. Hence, a scheduling policy must be formulated for allocating spectrum to \textit{appropriate} MS for establishing wireless communication links with \textit{appropriate} receivers at \textit{appropriate} transmit power.       
\begin{defi}[Policy for Resource Allocation and Power Control]
A policy $\Delta$ is an algorithm that chooses a power vector $\bl^\Delta(n)\in\cL$ in every slot $n$.
\end{defi}
Note that the decisions made by a policy can be based on the entire past and future information. 

Next, we will provide the details on the notion of \textit{rate stable} by relating the scheduling policies to the queue level dynamics and the average power expended. We will first define the average power consumed by MS $i$ under $\Delta$:
\begin{align}
\label{eq:avgPD}
\oP_i^\Delta = \lim\sup_{\10i T\to\infty}\frac{1}{T}\sum_{n=1}^T \sum_{\substack{j=0\\ j \neq i}}^N\l_{ij}^\Delta(n),
\end{align}
where $\l_{ij}^\Delta(n)$ is the transmit power on link $(i,j)$ under policy $\Delta$ in slot $n$. Define the average transmission rate on link $(i,j)$: 
\begin{align}
\label{eq:atrD}
\overline{\mu}_{ij}^\Delta = \lim\inf_{\10i T\to\infty}\frac{1}{T}\sum_{n=1}^T \mu_{ij}^\Delta(n),
\end{align} 
where $\mu_{ij}^\Delta(n)$ is compact notation for $\mu_{ij}(s(n),\bl^{\Delta}(n))$. Recall that when an MS sends data to the BS, i.e. on link $(i,0)$, it can either transmit packets backlogged in the \textit{own queue} or in the \textit{relay queue}. Let $\mathbb{I}_i^\Delta(n)$ denote the indicator function such that $\bI_i^\Delta(n) = 1$ if MS $i$ transmits its own traffic to the BS in slot $n$ under $\Delta$, and 0 otherwise. Now, we have:
\begin{align}
\label{eq:serv1}
 \overline{\mu}_{iO}^\Delta &= \lim\inf_{\10i T\to\infty}\frac{1}{T} \sum_{n=1}^T \mu_{i0}^\Delta(n) \mathbb{I}_i^\Delta(n), \mbox{\quad and} \\
\label{eq:serv2}
 \overline{\mu}_{iR}^\Delta &= \lim\inf_{\10i T\to\infty}\frac{1}{T} \sum_{t=1}^T \mu_{i0}^\Delta(n) [1-\mathbb{I}_i^\Delta(n)].
\end{align}
$\overline{\mu}_{iO}^\Delta$ and $\overline{\mu}_{iR}^\Delta$ are the service rates for the \textit{own queue} and the \textit{relay queue} of MS $i$, respectively, under resource allocation policy $\Delta$. A precise definition of rate stability associated with a policy $\Delta$ for a given arrival rate vector $\bm{\lambda}$ is: 
\begin{defi}
\label{defi:policy1}
A policy $\Delta$ is {\it feasible} if $\oP_i^\Delta \le \oP_i$ w.p.~1 for every MS $i$. Moreover, $\Delta$ is {\it rate stable} for arrival rate vector $\bm{\lambda}$, if (a)~$\Delta$ is feasible, (b)~$\overline{\mu}^\Delta_{iO} + \sum_{j\in \sN, j\neq i} \omu_{ij}^\Delta \ge \lambda_i$ w.p.~1 $\forall$ $i$, and (c)~$\omu_{iR}^\Delta \ge \sum_{j\in\sN, j \neq i}\omu_{ji}^\Delta$ w.p.~1 $\forall$ $i$.
\end{defi}
Condition (b) states that the sum of average service rate for the packets in the \textit{own queue} of each MS to the BS and to every other MS is greater than the average exogenous arrival rate. Condition (c) implies that the average service rate for \textit{relay queue} at an MS is greater than the sum of average arrival rates for the traffic coming from the \textit{own queue} of other MSs. We quantify the performance of a policy by its stability region.  
\begin{defi}[Rate region]
\label{defi:sysrate}
Stability region $\mathcal{C}^\Delta$ of policy $\Delta$ is the set of arrival rate vectors $\bm{\lambda}$ for which $\Delta$ is rate stable. System stability region $\overline{\mathcal{C}}$ is $\cup_{\Delta} \mathcal{C}^\Delta$.  
\end{defi}

The set $\overline{\mathcal{C}}$ contains all arrival rate vectors for which there exist a rate stable policy $\Delta$. Clearly, larger the stability region, better the policy is. In this regard, we define the following.   
\begin{defi}[Throughput optimal policy]
Policy $\Delta^\*$ is throughput optimal if $\mathcal{C}^{\Delta^\*} = \overline{\mathcal{C}}$.
\end{defi}
Next, we will quantify the system stability region.

\section{Quantification of System Stability Region}
\label{sec:QuantSysRateReg}
The generalized random walk of each MS collectively implies that network topology also evolves as generalized random walk. Our target is to provide a policy for the network evolving under generalized random walk. Towards this end, we first characterize the stability region under the assumption that network topology instead evolves in an IID fashion. This assumption helps us in appropriating strong law of large numbers. Let $\overline{\mathcal{C}}_{IID}$ be the system stability region under the assumption that network topology evolves in an IID fashion. Assume that in each slot, topology $s \in \cS$ is selected at random according to the probability distribution $\pi_{s}$ which is also the stationary distribution for our original network with topology evolving as generalized random walk. Let $\Lambda_{IID}$ denote the set of arrival rate vectors $\bm{\lambda}$ such that there exists some randomized policy identified by constants $w_{sk}$ and $q_{isk}$ for $i \in [N]$, $s\in S$ and $k=1,...,M$ satisfying 
\begin{gather}
\label{eq:Quantlamb1}
w_{sk} \in [0,1] \text{ and } \sum_{k=1}^M w_{sk} \leq 1 , \\
\label{eq:Quantlamb2}
q_{isk} \in [0,1], \\
\label{eq:Quantlamb3}
\sum_{s\in\cS}\sum_{k=1}^M\pi_{s}w_{sk} \left[ \mu_{i0}(s,\l_{k})q_{isk}+\sum_{j\in [N], j \neq i}\mu_{ij}(s,\l_{k})\right]\geq \lambda_{i},
\end{gather}
\begin{gather}
\sum_{s\in\cS}\sum_{k=1}^M\pi_{s}(1-q_{isk})w_{sk}\mu_{i0}(s,\l_{k}) \quad \nonumber \\
\label{eq:Quantlamb4}
\qquad \hspace{2cm}\geq \sum_{s\in\cS}\sum_{k=1}^M\sum_{j\in [N], j \neq i}\pi_{s}w_{sk}\mu_{ji}(s,\l_{k}),\\
\label{eq:Quantlamb5}
\sum_{s\in\cS}\sum_{k=1}^M\pi_{s}w_{sk}\sum_{j\in[N], j \neq i}\l_{ijk}\leq \oP_{i}.  
\end{gather}
Intuitively, $w_{sk}$ is the probability that power vector $\bl_{k} \in \cL$ is selected given that the network topology is $s \in \cS$, $q_{isk}$ is the probability that MS $i$ opts to transmit the packets in the \textit{own queue} to BS given that power vector $\bl_{k} \in \cL$ is selected and the network topology is $s \in \cS$. $\l_{ijk}$ is the power level of the link $(i,j)$, if it is active under the power vector $\bl_{k}$. Equation~(\ref{eq:Quantlamb3}) and (\ref{eq:Quantlamb4}) says that under the randomised policy identified by $w_{sk}$ and $q_{isk}$ for $i \in [N]$, $s\in S$ and $k=1,...,M$, average departure rate from the \textit{own queue} and \textit{relay queue} of MS $i$, respectively, should be greater than their respective average arrival rate.  Meanwhile, (\ref{eq:Quantlamb5}) guarantees that average power expended by the randomised policy satisfies the average power constraint. Then, we have the following.
\begin{theo}\label{thm:stab_region}
 The system stability region $\occ_{IID} = \Lambda_{IID}$. 
\end{theo}
The proof involves showing $\occ_{IID}\subseteq\Lambda_{IID}$ and $\Lambda_{IID}\subseteq\occ_{IID}$. For the former, we take a $\bm{\lambda}\in\occ_{IID}$ satisfying Definition~\ref{defi:policy1} and then show that there exists $w_{sk}$ and $q_{isk}$ for $i \in [N]$, $s\in S$ and $k=1,...,M$ which are fulfilling constraints (\ref{eq:Quantlamb1}) and (\ref{eq:Quantlamb2}), for which $\bm{\lambda}$ satisfies constraints (\ref{eq:Quantlamb3}), (11) and (\ref{eq:Quantlamb5}). The latter involves defining a Lyapunov function and then using it to show that for each $\bm{\lambda}\in\Lambda_{IID}$, there exists a randomized policy which is rate stable for that $\bm{\lambda}$, which by Definition~\ref{defi:sysrate} implies that  $\bm{\lambda}\in\occ_{IID}$. A rigorous analysis is given in \cite[Appendix B]{FV}. Note that the above characterization of system stability region  provides a recipe for a throughput optimal randomized policy $\Delta_{IID}$.

Taking into consideration of the sizes of current or future wireless networks, we would like to remark that it is not feasible to implement the randomized policy $\Delta_{IID}$. They are computationally expensive as we would need to determine appropriate probability distributions. The search for probability distribution requires one to know all system parameters in advance, which is impossible to determine unless we seek to know up to a bounded time in future. So, we desire an online scheduling policy. Nevertheless, we would utilize the randomized policy to investigate the throughput optimality of the online policy.

\section{Online Policy for Achieving System Stability}
\label{sec:OnThroPol}
In this section, we will formulate an online policy for resource allocation and policy control for network topology evolving as a generalized random walk. Afterwards, we will investigate its throughput optimality. Our process for formulating such a policy follows in three stages:
\begin{enumerate}
    \item Construct an online throughput optimal policy for network topology evolving in IID fashion.
    \item Construct an online $T-$step policy for similar topology evolution that uses stale information on queue lengths.
    \item Construct an online $T-$step policy for network topology evolving as generalized random walk that uses stale information on queue lengths.
\end{enumerate}
As a first step towards our goal, assume that network topology evolves in IID fashion. Consider a policy $\Delta_{IID}^{o}$ that solves the following optimization to obtain the power vector $\bl^{\Delta_{IID}^{o}}(n)$ in slot $n\geq 0$. 
\begin{align}
\label{eq:Maxopt}
&\textbf{Maximize: }  W(\textbf{X}(n),\textbf{Y}(n),\textbf{U}(n),s_{IID}(n),\bl^{\Delta_{IID}^{o}}(n)) = \nonumber \\ &\hspace{0.6cm}\sum_{i\in\sN}\Big\{\sum_{j\in\sN, j\neq i}[X_{i}(n)-Y_{j}(n)]\mu_{ij}(s_{IID}(n),\bl^{\Delta_{IID}^{o}}(n))\nonumber \\
&\hspace{1cm}+\max\{X_{i}(n),Y_i(n)\}\mu_{i0}(s_{IID}(n),\bl^{\Delta_{IID}^{o}}(n))\Big\}\nonumber \\
&\hspace{1cm}-\sum_{i\in\sN}U_{i}(n)\sum_{j=0, j \neq i}^N\l_{ij}^{\Delta_{IID}^{o}}(n) \nonumber \\
&\textbf{Subject to: }\bl^{\Delta_{IID}^{o}}(n)\in\cL.
\end{align}
where $s_{IID}(n)$ is the network topology in slot $n$. The objective function can be interpreted as follows: inside the summation over $i$, the first term is  summation of products of back-pressure involving own queue at MS $i$ w.r.t the relay queue at an MS $j$ and the transmission rate of link $(i,j)$. The second term is the product of the maximum of the back-pressure of \textit{own queue} and \textit{relay queue} with respect to BS and the link connecting MS with the BS. The third term is the product of instantaneous \textit{virtual queue} length with arrivals in that queue. The foloowing lemma states that $\mathcal{C}_{IID}^{\Delta_{IID}^{o}} = \occ_{IID} = \Lambda_{IID}$.

\begin{lem}
\label{lem:OptIID}
Let $\bm{\lambda}$ be any stabilizable arrival rate vector with network topology changing in IID fashion in each slot. Then, $\Delta_{IID}^{o}$ stabilizes the system.
\end{lem}
\begin{IEEEproof}
We provide a brief outline of the proof. For detailed proof, refer to \cite[Appendix C]{FV}. We use the general framework of Lyapunov optimization introduced in \cite{TE} to prove stability. Towards that end, we define the following Lyapunov function
\begin{align}
\label{eq:Lyaopt}
    f(n) = \sum_{i \in [N]}\bigg[(X_{i}(n))^{2} + (Y_{i}(n))^{2} + (U_{i}(n))^{2}\bigg]. 
\end{align}
Our goal is to determine the set $\mathcal{A}$ of vectors $\{\mathbf{X}(n), \mathbf{Y}(n), \mathbf{U}(n)\}$ such that the expected Lyapunov drift in any slot $n \geq 0$ outside $\mathcal{A}$ is negative. That is,
\begin{align*}
    \mathop{\mathbb{E}}& [f(n+1)-f(n)|\mathbf{X}(n),\mathbf{Y}(n),\mathbf{U}(n)] \nonumber \\
    &< \begin{cases}
        \infty, & \forall \text{ } \{\mathbf{X}(n),\mathbf{Y}(n),\mathbf{U}(n)\} \\
        -1, & \text{if } \{\mathbf{X}(n),\mathbf{Y}(n),\mathbf{U}(n)\} \notin \mathcal{A}. 
  \end{cases}
\end{align*}
Using the fact that for every randomized policy $\Delta_{IID}$ and $n\geq0$,
\begin{align}
    &W(\textbf{X}(n),\textbf{Y}(n),\textbf{U}(n),s_{IID}(n),\bl^{\Delta_{IID}^{o}}(n)) \nonumber \\
    &  \hspace{0.6cm}\geq W(\textbf{X}(n),\textbf{Y}(n),\textbf{U}(n),s_{IID}(n),\bl^{\Delta_{IID}}(n)), \nonumber
\end{align}
we show that $\mathcal{A}$ is finite for every $\bm{\lambda} \in \occ_{IID}$. Then, the Foster's Theorem  \cite[Theorem~2.2.3]{GVM} guarantees that $\{\mathbf{X}(n), \mathbf{Y}(n), \mathbf{U}(n)\}_{n \geq 0}$ is positive recurrent, and for each queue the expected queue length under its stationary distribution is finite. Thus, the system is stable under $\Delta_{IID}^{o}$ $\forall \bm{\lambda} \in \Lambda$.
\end{IEEEproof}

Next, consider that the time axis is divided in the intervals of length $T$, i.e., in intervals of the form $[KT,(K+1)T-1]$. Under the assumption that network topology evolves in IID fashion, we develop a policy $\Delta_{IID}^{T}$ that solves the following optimization to obtain the power vector $\bl^{\Delta_{IID}^{T}}(n)$ in each slot $n \in [KT,(K+1)T-1]$. 
\begin{align}
\label{eq:MaxOptTstepIID}
&\textbf{Maximize:} \nonumber \\ 
& W(\textbf{X}(KT),\textbf{Y}(KT),\textbf{U}(KT),s_{IID}(n),\bl^{\Delta_{IID}^{T}}(n)) =\nonumber \\
&\hspace{0.1cm}\sum_{i\in\sN}\Big\{\sum_{j\in\sN, j\neq i}[X_{i}(KT)-Y_{j}(KT)]\mu_{ij}(s_{IID}(n),\bl^{\Delta_{IID}^{T}}(n)\nonumber \\
&\hspace{1cm}+\max\{X_{i}(KT),Y_i(KT)\}\mu_{i0}(s_{IID}(n),\bl^{\Delta_{IID}^{T}}(n)\Big\}\nonumber \\
&\hspace{1cm}-\sum_{i\in\sN}U_{i}(KT)\sum_{j=0, j \neq i}^N\l_{ij}^{\Delta_{IID}^{T}}(n) \nonumber \\
&\textbf{Subject to: }\bl^{\Delta_{IID}^{T}}(n)\in\cL. 
\end{align}
The policy $\Delta_{IID}^{T}$ computes $\bl^{\Delta_{IID}^{T}}(n)$ based on stale information on queue lengths from the beginning of the interval, i.e., in the slots $KT$ for $K \geq 0$. The information on the queue lengths in the optimization are updated at the beginning of each interval. The following lemma shows that even with stale updates in queue lengths, one can obtain $\mathcal{C}_{IID}^{\Delta_{IID}^{T}} = \occ_{IID}$.

\begin{lem}
\label{lem:Tstepstable}
Let $\bm{\lambda}$ be any stabilizable arrival rate vector with system topology changing in IID fashion in each slot. Then, $\Delta_{IID}^{T}$ stabilizes the system.
\end{lem}
\begin{IEEEproof}
We use similar steps in the proof as that of Lemma~\ref{lem:OptIID}. The crucial step in showing the negative Lyapunov drift outside some set $\mathcal{B}$ of vectors $\{\mathbf{X}(KT),\mathbf{Y}(KT),\mathbf{U}(KT)\}$ is
\begin{align}
        &W(\mathbf{X}(n),\mathbf{Y}(n),\mathbf{U}(n),s_{IID}(n),\bl^{\Delta_{IID}^{o}}(n))\nonumber  \\
        &\leq W(\mathbf{X}(KT),\mathbf{Y}(KT),\mathbf{U}(KT),s_{IID}(n),\bl^{\Delta_{IID}^{T}}(n))  \nonumber \\ & \hspace{3cm} + \frac{M_{2}}{2}, \nonumber 
\end{align}
$\forall$ $n \in [KT, (K+1)T-1]$ and for some constant $M_{2}$. For the detailed proof, see \cite[Appendix D]{FV}.
\end{IEEEproof}

We finally formulate an online policy $\Delta_{RW}^{T}$ for our original network topology where network evolves as a generalized random walk. Again, consider the time axis to be divided in the intervals of length $T$, i.e., $[KT,(K+1)T-1]$.  Consider the $s_{RW}(n)$ be the network topology when the network evolves as a generalized random walk. Our policy $\Delta_{RW}^{T}$ solves the following optimization to obtain the power vector $\bl^{\Delta_{RW}^{T}}(n)$ in each slot $n \in [KT,(K+1)T-1]$.
\begin{align}
\label{eq:MaxOptTstepRW}
&\textbf{Maximize:}\nonumber \\
& W(\mathbf{X}(KT),\mathbf{Y}(KT),\mathbf{U}(KT),s_{RW}(n),\bl^{\Delta_{RW}^{T}}(n))\nonumber \\
&\hspace{0.1cm}=\sum_{i\in\sN}\Big\{\sum_{\substack{j\in\sN\\ j\neq i}}[X_{i}(KT)-Y_{j}(KT)]\mu_{ij}(s_{RW}(n),\bl^{\Delta_{RW}^{T}}(n)\nonumber \\
&\hspace{1cm}+\max\{X_{i}(KT),Y_i(KT)\}\mu_{i0}(s_{RW}(n),\bl^{\Delta_{RW}^{T}}(n)\Big\}\nonumber \\
&\hspace{1cm}-\sum_{i\in\sN}U_{i}(KT)\sum_{j=0, j \neq i}^N\l_{ij}^{\Delta_{RW}^{T}}(n) \nonumber \\
&\textbf{Subject to: }\bl^{\Delta_{RW}^{T}}(n)\in\cL. 
\end{align}
Consider $J$ is a $1\times N$-dimensional vector with all entries equal to $1$. We have the following theorem to comment upon its throughput optimality.
\begin{theo}\label{thm:RWoptimal}
For some $\epsilon > 0$, define $\Lambda_{\epsilon} \myeq \{\bm{\lambda} \mid \bm{\lambda} + \epsilon J \in \Lambda_{IID}\}$. Then, for any $T>0$, there exists some constant $c$ such that $\forall \epsilon > c$, $\Delta_{IID}^{T}$ stabilizes the system for all such $\bm{\lambda} \in \Lambda_{\epsilon}$.
\end{theo}
\begin{IEEEproof}
We provide the proof outline. Consider the same Lyapunov function as defined in (\ref{eq:Lyaopt}). We consider two similar networks but one network having topology evolving in IID fashion and the other network having topology evolving as generalized random walk. Assume that the stationary distribution of the network with topology evolving as generalized random walk is same as IID distribution of the first network. Let $s_{IID}(n)$ and $s_{RW}(n)$ be the state of the grid under IID and Markov mobility process at time $n$, respectively. Suppose that $s_{IID}(n) = s_{RW}(n)$ for some $n \in [KT, (K+1)T-1]$. Then, we have
\begin{align}
    &W(\mathbf{X}(KT),\mathbf{Y}(KT),\mathbf{U}(KT),s_{IID}(n),\bl^{\Delta_{IID}^{T}}(n)) \nonumber \\
    \label{eq:Wcomp2}
    &= W(\mathbf{X}(KT),\mathbf{Y}(KT),\mathbf{U}(KT),s_{RW}(n),\bl^{\Delta_{RW}^{T}}(n)). 
\end{align}
Also, we can show that for  $s_{IID}(n) \neq s_{RW}(n)$ in some slot 
$n \in [KT, (K+1)T-1]$, we have
\begin{align}
       &W(\mathbf{X}(KT),\mathbf{Y}(KT),\mathbf{U}(KT),s_{IID}(n),\bl^{\Delta_{IID}^{T}}(n)) \nonumber \\ 
       &\leq W(\mathbf{X}(KT),\mathbf{Y}(KT),\mathbf{U}(KT),s_{RW}(n),\bl^{\Delta_{RW}^{T}}(n)) \nonumber \\
       &+ \mu_{max}(N+1)\bigg(\sum_{i \in [N]}X_{i}(KT) + \sum_{i \in [N]}Y_{i}(KT)\Bigg) \nonumber \\
       \label{eq:Wcomp3}
       &+ P_{max}\sum_{i \in N}U_{i}(KT).
\end{align}
For tractability, suppose there are two possible states, $s_1$ and $s_2$, that network topology can assume. Let 
\begin{equation}
\begin{split}
    \tau = & \mid \text{number of states } s_1 \text{ in IID process } \\
    &- \text{number of states } s_1 \text{ in Markov process}\mid. \nonumber
\end{split}
\end{equation}
Suppose $\epsilon >  \frac{1}{2}\max\{(N+1)\mu_{max}\mathop{\mathbb{E}}[\tau],P_{max}\mathop{\mathbb{E}}[\tau]\}$. Then, using (\ref{eq:Wcomp2}) and (\ref{eq:Wcomp3}), we can show that the Lyapunov drift is negative outside a finite set $\mathcal{C}_{\epsilon}$ of vectors $\{\mathbf{X}(KT),\mathbf{Y}(KT),\mathbf{U}(KT)\}$ $\forall \bm{\lambda} \in \Lambda_{\epsilon}$. Then, we apply Foster's Theorem to conclude that system is stable under $\Delta_{IID}^{T}$ $\forall \bm{\lambda} \in \Lambda_{\epsilon}$. It is important to note that due to exponentially fast convergence of empirical distribution to the unique stationary distribution for ergodic Markov chains \cite{AA}, we have $\mathop{\mathbb{E}}[\tau] \rightarrow 0$ as $T \rightarrow \infty$ and, thus, $\Lambda \setminus \Lambda_{\epsilon} \rightarrow \O$. For the detailed proof, see \cite[Appendix E]{FV}.
\end{IEEEproof}
We term such policies that stabilize the system for arrival rate vectors $\bm{\lambda} \in \Lambda_{\epsilon}$ as $\epsilon-$throughput optimal.

\section{Numerical Results}
\label{sec:NumRes}
In this section, we evaluate the performance of the proposed throughput optimal policy $\Delta_{RW}^{T}$ for resource allocation and power control in a D$2$D relay-assisted cellular network. Towards this end, we simulated a $2$-D grid with vertices of the form $(a,b)$ such that $a \in \{0, 10, 20, ..., 2000\}$,$b \in \{0, 10, 20, ..., 2000\}$ in meters. Thus, two adjacent grid points are $10$ meters away from each other. The MSs are restricted to move along the grid. Assume that the BS is located at $(1000,1000)$. In this simulation, $60$ MSs are present in the grid. These MSs move in a reflected random walk fashion with each MS in each slot equally likely to go to any of its neighbouring grid point or stay at the same grid point. For convenience, assume that when any two MSs in the same grid point, they are separated by a distance of $1$ meter. Each MS can assume power level from the set $\mathcal{P} = \{20, 23.01, 24.77, 26.02, 26.99, 27.78, 28.45, 29.03, 29.54 ,30\}$ (in dBm) in each slot. The average power constraint at each MS is $28$ dBm. The carrier frequency is taken to be $3.5$ GHz. Applying OFDMA, there are $50$ physical resource blocks (PRBs) to be assigned to various links amounting to total bandwidth of  $10$ MHz. For simplicity in simulation, it is assumed that each PRB can be assigned to only one MS in each slot, thereby, effectively removing interference. Thus, the transmission rate of a link is dependent only on the power level of the transmitting MS of that link. Thus, for any link $(i,j)$, $\mu_{ij}(\bs_{RW}(n),\bl^{\Delta_{IID}^{T}}(n)) = \mu_{ij}(\bs_{RW}(n),\l^{\Delta_{IID}^{T}}_{ij}(n))$ However, we emphasize that that the theory developed in this work applies for any general interference scenario. 

\begin{algorithm}[ht]
  \SetKwData{Left}{left}\SetKwData{This}{this}\SetKwData{Up}{up}
  \SetKwFunction{FMain}{OptPow}
  \SetKwProg{Fn}{Function}{:}{}
  \SetKwInOut{Input}{input}\SetKwInOut{Output}{output}
  \Input{$MS\_list:$ List of ID of users\\
  $PRB\_list:$ List of ID of PRBs\\
  $MS\_loc:$ Location of MSs\\
  $BS\_loc:$ Location of BS\\ 
  $\mathcal{P}:$ List of power levels}
  \Output{Assignment of PRBs with appropriate power} 
  Weighted Bipartite graph $\mathcal{G} = \phi$ \;
  $Rx\_list = \phi$ \;
  \For{$prb \in PRB\_list$}  {
  	\For{$Tx \in MS\_list$}{
		\For {Rx = BS}{
			$\Gamma$($Tx,Rx,prb$) = OptPow($Tx\_loc$, $MS\_loc$, $BS\_loc$, $\mathcal{P}$, $0$)\;
		}
		\For {Rx $\in$ MS\_list-\{Tx\} }{
			$\Gamma$($Tx,Rx,prb$) = OptPow($Tx\_loc$, $MS\_loc$, $BS\_loc$, $\mathcal{P}$, $1$)\;
		}
		$edge = \{Tx,prb\}$\;
		$weight(\mathcal{G}(edge)) = \max_{Rx \in \{BS\} \cup  MS\_list-\{Tx\}} \Gamma(Tx,Rx,prb)$\;
		$Rx\_list(Tx) = \arg\max_{Rx \in \{BS\} \cup  MS\_list-\{Tx\}} \Gamma(Tx,Rx,prb)$\;	
	}
  }
  
  Perform Maximum Bipartite Matching of $\mathcal{G}$\;
  
\caption{$\epsilon-$throughput optimal policy}\label{algo1}
\end{algorithm}

\begin{algorithm}[ht]
  \DontPrintSemicolon
  \SetKwFunction{FMain}{OptPow}
  \SetKwProg{Fn}{Function}{:}{}
  \Fn{\FMain{$Tx\_loc$, $MS\_loc$, $BS\_loc$, $\mathcal{P}$,$flag$}}{
  	\eIf{$flag=0$}{
		Find $P \in \mathcal{P}$ that maximises \\
		$\gamma = \max[X_{Tx}(n),Y_{Tx}(n)]\times\mu_{Tx,BS}(\bs_{RW}(n),P)- U_{Tx}(n)\times P$\;	
	}{
		Find $P \in \mathcal{P}$ that maximises \\
		$\gamma = [X_{Tx}(n) - Y_{Rx}(n)]\times \mu_{Tx,Rx}(\bs_{RW}(n),P)- U_{Tx}(n)\times P$	\;
	}
 
        \KwRet\ $\gamma$\;
  }
\caption{Function description of OptPow}\label{algo2} 
\end{algorithm}  

\begin{figure*}
    \centering
     \begin{subfigure}[b]{0.45\linewidth}
        \includegraphics[width=\linewidth]{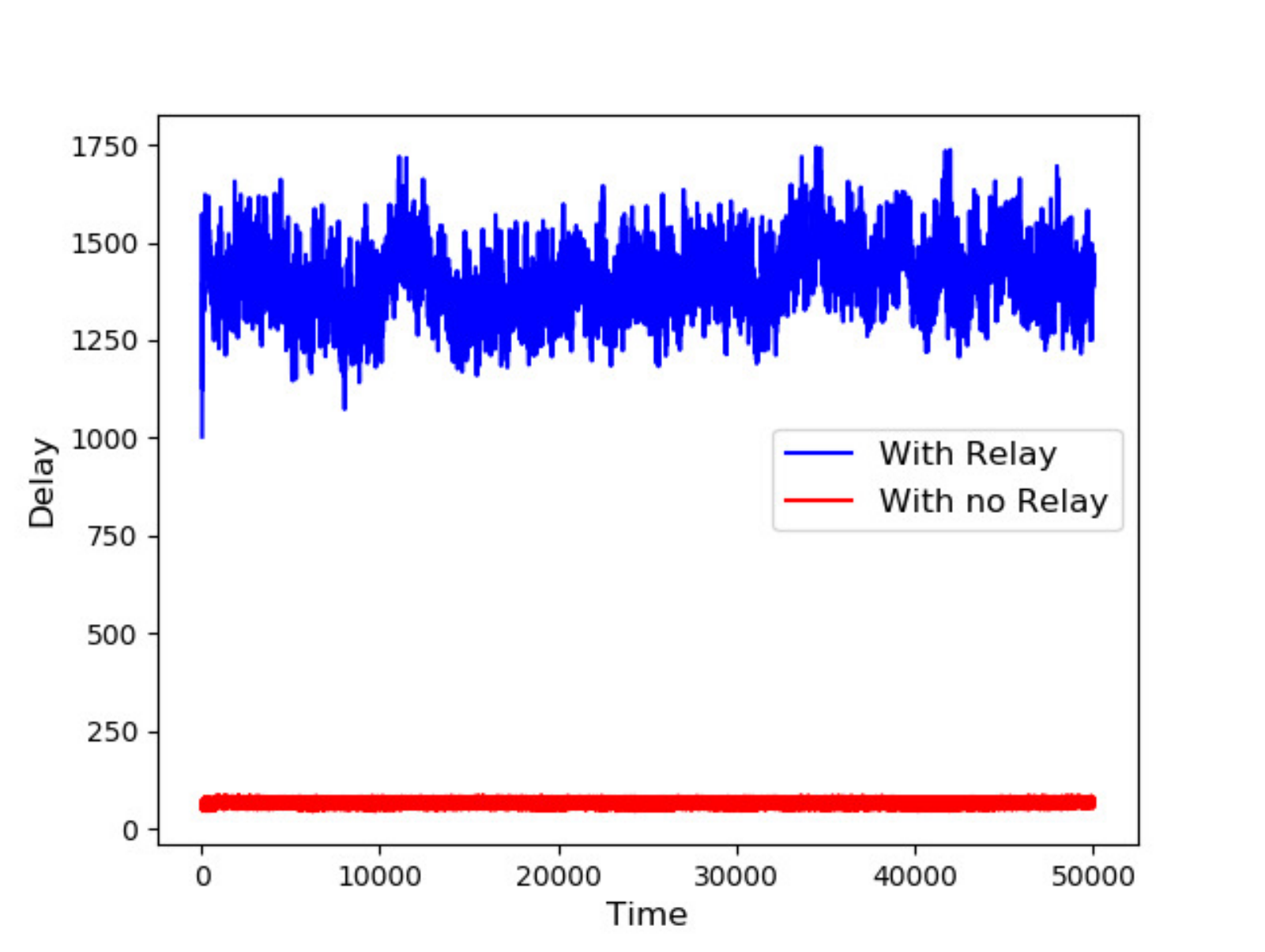}
        \caption{Arrival Rate = 18 packets/slot}
        \label{fig:arrival18}
    \end{subfigure}   
    \begin{subfigure}[b]{0.45\linewidth}
        \includegraphics[width=\linewidth]{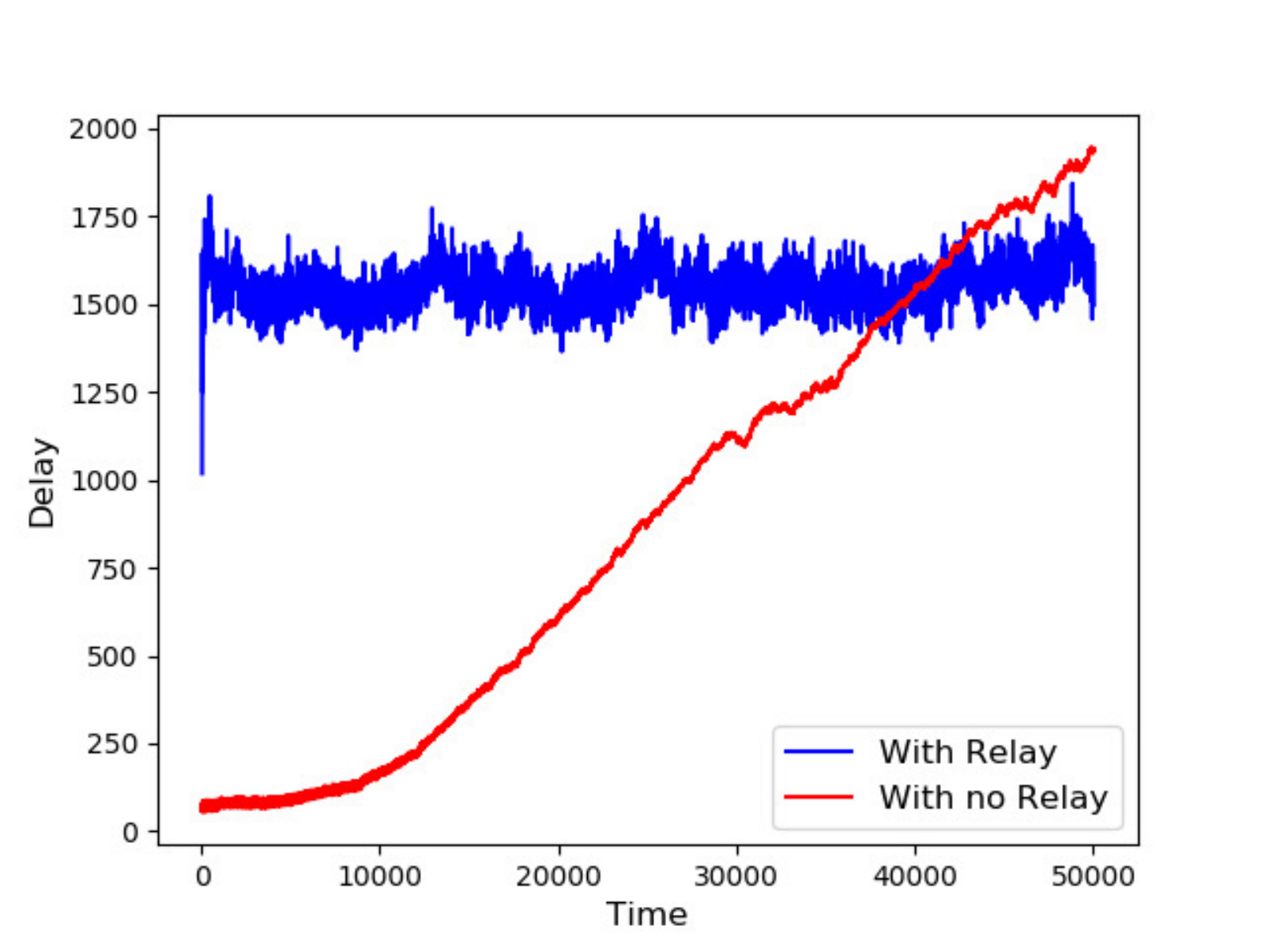}
        \caption{Arrival Rate = 20 packets/slot}
        \label{fig:arrival20}
    \end{subfigure}\\

    \begin{subfigure}[b]{0.45\linewidth}
        \includegraphics[width=\linewidth]{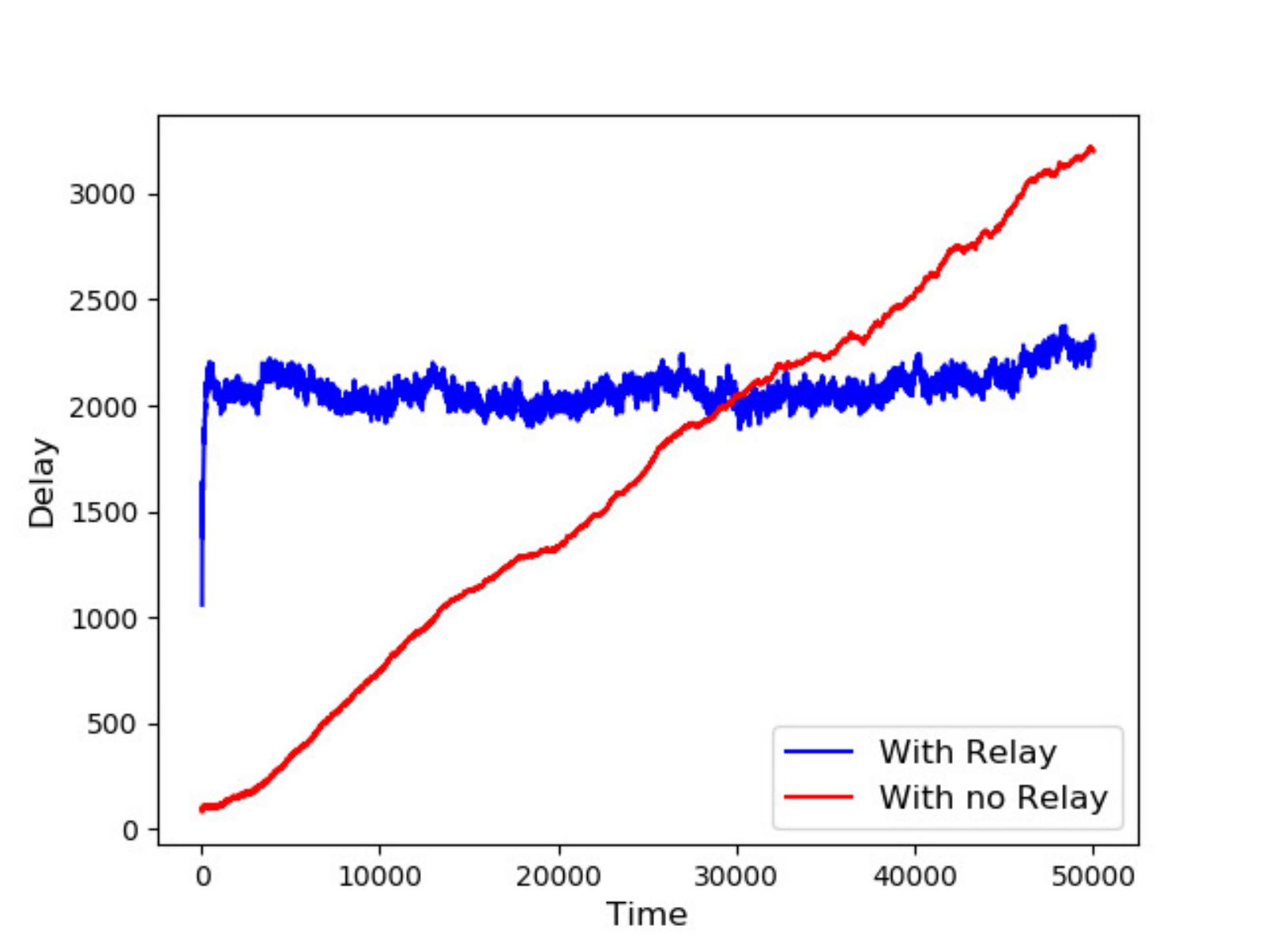}
        \caption{Arrival Rate = 25 packets/slot}
        \label{fig:arrival25}
    \end{subfigure}
    \begin{subfigure}[b]{0.45\linewidth}
        \includegraphics[width=\linewidth]{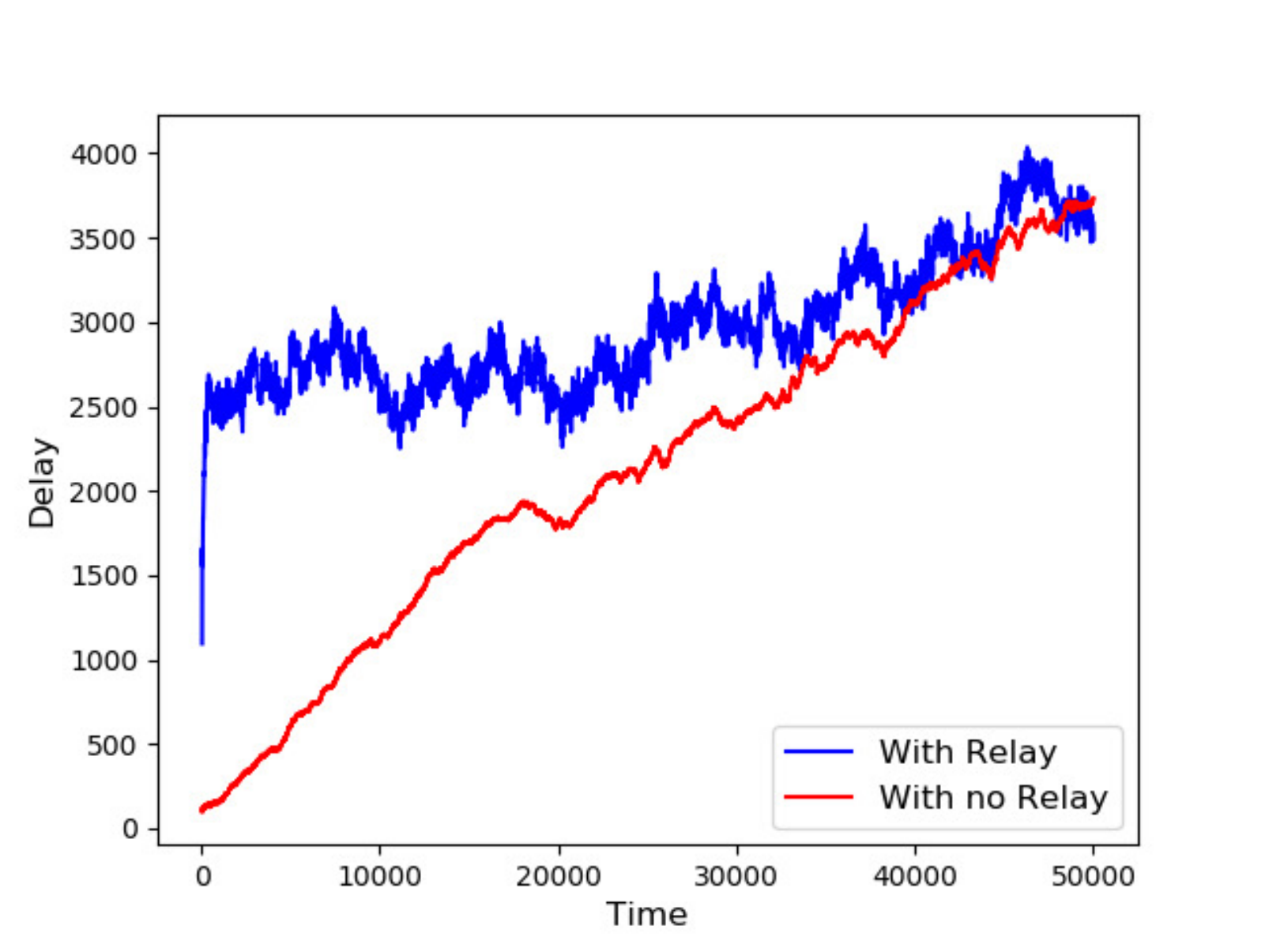}
        \caption{Arrival Rate = 28 packets/slot}
        \label{fig:arrival28}
    \end{subfigure}    
    ~ 
    \caption{Average Delay at each MS for various arrival rates. In each figure, the arrival process is a Poisson process with corresponding arrival rate. Note that we have used queue length and delay interchangeably.}\label{fig:Delay}
\end{figure*}

We evaluate the performance of our policy described in (\ref{eq:MaxOptTstepRW}) in two scenarios: (a) D$2$D relay-assisted cellular network as proposed in Section~\ref{sec:SysMod} and (b) cellular network with no relays. Note that in scenario (b), each MS just maintains \emph{own queue} that backlogs the packets to be sent directly to the BS. Thus, for scenario (b), (\ref{eq:MaxOptTstepRW}) modifies to
\begin{align}
\label{eq:Maxopt_no_relay}
\textbf{Maximize: }&\sum_{i\in\sN}\Big\{\sum_{\substack{j\in\sN\\ j \neq i}}X_{i}(KT)\mu_{i0}(\bs_{RW}(n),\l^{\Delta_{RW}^{T}}_{ij})\Big\}\nonumber \\
&-\sum_{i\in\sN}U_{i}(n)\sum_{\substack{j=0\\ j\neq i}}^N\l_{ij}^{\Delta_{RW}^{T}} \nonumber \\
\textbf{Subject to: }&\bl^{\Delta_{RW}^{T}}\in\cL. 
\end{align}
For evaluation of transmission rate $\mu_{Tx,Rx}(s_{RW}(n),\bl^{\Delta_{RW}^{T}})$, we use alpha-beta-gamma (ABG) path loss model for $5$G urban micro and macro-cellular scenarios \cite{STS}. For scaling purpose, assume that $1$ packet = $25000$ bits. For brevity, we assume that all MSs have symmetrical arrival rate. Note that we have used delay and queue length interchangeably. Fig.~\ref{fig:Delay} illustrates the average delay per node for symmetric arrival rates of ${18,20,25,28}$ packets/slot. Each of these simulations were run for $50000$ slots. It is important to note that for the scenario where relays are present, we have to calculate the sum of queue lengths in \emph{own queue} and \emph{relay queue} whereas in the scenario where relays are absent, we have to calculate queue length in \emph{own queue} only. Clearly, from Fig.\ref{fig:Delay}, in both scenarios (a) and (b), the average queue length increases with arrival rate. Furthermore, it can be inferred that for the scenario (b), where relays are absent, the maximum stabilizable symmetric arrival rate lies between $18$ packets/slot and $20$ packets/slot while for scenario (a), where relays are present, the maximum stabilizable symmetric arrival rate lies between $25$ packets/slot and $28$ packets/slot. Thus, using relay increases the maximum symmetric stabilizable arrival rate by at least $25\%$. This implies that using mobile relays can be beneficial in alleviating the detrimental impact of mobility on delay \cite{AJBD}.

To implement proposed $\epsilon-$throughput optimal policy, we refer to Algorithm~\ref{algo1}. We construct a weighted bipartite graph where the end-points of an edge represent a particular MS and a particular PRB.  The edge symbolises assignment of the PRB to the MS. For finding the weight of an edge between a MS, say $i$, and a PRB, say $prb$, first evaluate the value of the following weights by calling Algorithm~\ref{algo2}: 
\begin{align}
&1. \hspace{0.3cm} \text{For a link from MS $i$ to any other MS $j$, obtain  }  \nonumber \\
&\max_{P \in \mathcal{P}}\bigg\{[X_{i}(n) - Y_{j}(n)]\mu_{ij}(\bs_{RW}(n),P)- U_{i}(n) P \bigg\}. \nonumber
\end{align}
\begin{align}
&2. \hspace{0.3cm} \text{For a link from MS $i$ to the BS, obtain }  \nonumber \\
&\max_{P \in \mathcal{P}}\bigg\{max[X_{i}(n),Y_{i}(n)] \mu_{i,BS}(\bs_{RW}(n),P)- U_{i}(n) P\bigg\}. \nonumber
\end{align}
Then, determine the link from MS $i$ that gives highest weight. The corresponding weight is the weight of the edge between MS $i$ and PRB $prb$. One can apply suitable algorithm to find the maximum weighted bipartite matching of the above constructed bipartite graph to obtain an assignment of PRBs to the MSs such that objective in (\ref{eq:MaxOptTstepRW}) is satisfied. The Python implementation for Algorithm~\ref{algo1} is provided in \cite{CODE}.

\section{Conclusions}
\label{sec:Con}
In this paper, we investigated resource allocation and power control for delay-sensitive applications in D$2$D relay-assisted cellular communication. We evaluated the system stability region when the network topology evolves in IID fashion. Thereafter, we formulated an online $T-$step policy for network topology evolving in a generalized random walk that is shown to be $\epsilon-$throughput optimal with respect to above system stability region. We have also provided numerical results to illustrate the benefits of having relays. The future work will involve formulating a distributed policy for resource allocation and power control which is necessary from the point of scalability and robustness.

\newpage

\end{document}


\title{Stability Analysis of Device-to-Device Relay-Assisted Cellular Networks: Proofs}

\remove{
\author{\IEEEauthorblockN{Under Construction}
\IEEEauthorblockA{Electrical Engineering Department \\
Indian Institute of Technology Bombay\\
Mumbai, India 400 076\\
Email: Under Construction}}}

\author{Soubhik Deb, Prasanna Chaporkar,  and Abhay Karandikar %
}

\maketitle

\appendices
\section{Fluid Limits} 
In this section, we present the basic concepts on fluid limits for the queues in our system and its associated lemmas with respect to our system model. For further study, kindly refer to \cite{DAI}. These lemmas are crucial in proving \cite[Theorem 1]{ICC}.

Before, we proceed further, we introduce some notations. \cite[Equation 2]{ICC}, \cite[Equation 3]{ICC} and \cite[Equation 4]{ICC} can be respectively rephrased as follows:
\begin{align}
\label{eq:X_cumu}
X_{i}(n)&=X_{i}(0)+\cA_i(n)-\sum_{j\in\sN, j\neq i}\cD_{ij}(n)-\cD_{i0}(n),\\
\label{eq:Y_cumu}
Y_{i}(n)&=Y_{i}(0)+\sum_{j\in\sN, j \neq i}\cD_{ji}(n)-\cD_{iR}(n), \\
\label{eq:U_cumu}
U_{i}(n)& = U_{i}(0) + \cA_{Ui}(n) - \cD_{Ui}(n),
\end{align}
where $\cA_{i}(n)$ is the total exogenous arrivals  in MS $i$ till slot $n$, $\cD_{ij}(n)$ is the total number of packets transmitted by MS $i$ to MS $j$ until slot $n$, $\cD_{i0}(n)$ ($\cD_{iR}(n)$) denotes the number of departures to BS from the \textit{own} (\textit{relay}) \textit{queue} until slot $n$. $\cA_{Ui}(n)$ and $\cD_{Ui}(n)$ be the total arrivals and departures, respectively, that takes place in the \textit{virtual queue} of MS $i$ till slot $n$.

Suppose the set of non-negative integers and reals are denoted by $\mathbb{N}$ and $\mathbb{R}$, respectively. The domain of the functions $X_{i}(n), Y_{i}(n), \cA_{i}(n), \cD_{iO}(n), \cD_{iR}(n), \cD_{ij}(n), U_{i}(n), \cA_{Ui}(n), \\ \cD_{Ui}(n)$ is $\mathbb{N}$. We define these functions for arbitrary $t \in \mathbb{R}$ by using a piecewise linear interpolation. The piecewise linear interpolation of a function is defined as: for any function $f$ with $\mathbb{N}$ as its domain, $t\in (n,n+1]$, 
\begin{align}
f(t)=f(n)+(t-n)(f(n+1)-f(n)).  \nonumber
\end{align}
Suppose that for any scheduling policy, the number of packets that can be transmitted across any link in a slot is bounded by $\mu_{max}$. Also, as mentioned before, the maximum number of exogenous arrivals at each MS $i$ in a slot is bounded by $A_{max}$. Further, each MS $i$ can utilize at most $P_{max}$ power in each slot. For a random process ${f(t)}_{t\geq 0}$, we show the randomness of the sampling path $\omega$ by denoting its value at time $t$ along the sample path $\omega$ by $f(t,\omega)$. For every $\omega$, $\forall i \in [N]$ and $t\geq 0$, we can easily show that $X_{i}(t,\omega),Y_{i}(t,\omega),\cA_{i}(t,\omega),\cD_{iO}(t,\omega),\cD_{iR}(t,\omega),\cD_{ij}(t,\omega),\\U_{i}(t,\omega),\cA_{Ui}(t,\omega),\cD_{Ui}(t,\omega)$ are all Lipschitz functions. 
\remove{\begin{align}
&X_{i}(t+\delta,\omega)-X_{i}(t,\omega)\leq A_{max}\delta \nonumber \\
&Y_{i}(t+\delta,\omega)-Y_{i}(t,\omega)\leq N\mu_{max}\delta \nonumber \\
&A_{i}(t+\delta,\omega)-A_{i}(t,\omega)\leq A_{max}\delta\nonumber \\
&D_{iD}(t+\delta,\omega)-D_{iD}(t,\omega) \leq \mu_{max}\delta \nonumber \\
&D_{iR}(t+\delta,\omega)-D_{iR}(t,\omega) \leq \mu_{max}\delta \nonumber \\
&D_{ij}(t+\delta,\omega)-D_{ij}(t,\omega) \leq \mu_{max}\delta \nonumber \\
&U_{i}(t+\delta,\omega)-U_{i}(t,\omega)\leq P_{max}\delta \nonumber \\
&A_{Ui}(t+\delta,\omega)-A_{Ui}(t,\omega)\leq P_{max}\delta\nonumber \\
&D_{Ui}(t+\delta,\omega)-D_{Ui}(t,\omega) \leq \oP_{i}\delta \nonumber
\end{align} }
Now, let us define fluid scaling of any given functions $f(.)$ as follows:
\begin{align}
f^{r}(t,\omega)\myeq\frac{f(rt,\omega)}{r}, \hspace{5mm}\text{for every $r > 0$}. \nonumber
\end{align}
It follows that for every $r > 0$
\begin{align}
&X_{i}^{r}(t+\delta,\omega)-X_{i}^{r}(t,\omega)\leq A_{max}\delta. \nonumber 
\end{align}
We have similar bounds on $Y_{i}^{r}(t,\omega), \cA_{i}^{r}(t,\omega), \cD_{iO}^{r}(t,\omega), \cD_{iR}^{r}(t,\omega), \cD_{ij}^{r}(t,\omega), U_{i}^{r}(t,\omega), \\ \cA_{Ui}^{r}(t,\omega), \cD_{Ui}^{r}(t,\omega)$. 

Thus, all the above functions are Lipschitz continuous, and hence uniformly continuous on any compact interval. Clearly, the above functions are also bounded on any compact interval. Fix a compact interval $[0,t]$. Now, consider any sequence $r_{n}$ such that $r_{n} \to \infty$ as $n \to \infty$. Then, by Arzela-Ascoli theorem \cite{WR}, there exists a subsequence $r_{n_{k}}$ and continuous functions $\widetilde{X}_{i}(.)$ such that for every $i,\omega$
\begin{align}
&\lim_{k\to\infty}\sup_{\hat{t}\in[0,t]}\vert X_{i}^{r_{n_{k}}}(\hat{t},\omega)-\widetilde{X}_{i}(\hat{t})\vert=0. \nonumber 
\end{align} 
Similar convergence results for other system processes exist. Let the functions to which $Y_{i}^{r}(t,\omega), \cA_{i}^{r}(t,\omega), \cD_{i0}^{r}(t,\omega), \cD_{iR}^{r}(t,\omega), \cD_{ij}^{r}(t,\omega), U_{i}^{r}(t,\omega),\\ \cA_{Ui}^{r}(t,\omega), \cD_{Ui}^{r}(t,\omega)$ converge, for some subsequence, be denoted by $\widetilde{Y}_{i}(t), \widetilde{\cA}_{i}(t), \widetilde{\cD}_{i0}(t), \widetilde{\cD}_{iR}(t), \widetilde{\cD}_{ij}(t), \widetilde{U}_{i}(t), \widetilde{\cA}_{Ui}(t),\\ \widetilde{\cD}_{Ui}(t)$, respectively. We will now define fluid limits.
\begin{defi}
\label{defi:definition5}	
$\widetilde{X}_{i}(.), \widetilde{Y}_{i}(.), \widetilde{\cA}_{i}(.),  \widetilde{\cD}_{i0}(.), \widetilde{\cD}_{iR}(.), \widetilde{\cD}_{ij}(.), \widetilde{U}_{i}(.),\\ \widetilde{\cA}_{Ui}(.), \widetilde{\cD}_{Ui}(.)$ are called fluid limits if there exists $r_{n_{k}}$ such that all the aforementioned convergence relations are satisfied.
\end{defi}
Now, we state some important properties of the fluid limits.
\begin{lem}
\label{lem:lem1}
Every fluid limit satisfies,  $\widetilde{\cA}_{i}(t)=\lambda_{i}t$ $w.p.$ 1 for every MS $i$ and $t\geq0$.
\end{lem} 
\begin{IEEEproof}
Since $\widetilde{\cA}_{i}(t)$ is a fluid limit, thus, by Definition~\ref{defi:definition5}, there exists a sequence $r_{n_{k}}$ such that $\lim_{k\to\infty}r_{n_{k}}=\infty$ and 
\begin{align}
\widetilde{\cA}_{i}(t)&=\lim_{k\to\infty}\cA_{i}^{r_{n_{k}}}(t) \nonumber \\
&=\lim_{k\to\infty}\frac{\cA_{i}(r_{n_{k}}t)}{r_{n_{k}}} \nonumber \\
&=\lim_{k\to\infty}\frac{\cA_{i}(r_{n_{k}}t)}{r_{n_{k}}t}t \nonumber \\
&=\lambda_{i}t \text{ w.p. 1 (by strong law of large numbers (SLLN))}. \nonumber
\end{align}
\end{IEEEproof}  
\begin{lem}
\label{lem:lem2}
Any fluid limit $\widetilde{X}_{i}(.),  \widetilde{\cA}_{i}(.),  \widetilde{\cD}_{i0}(.), \widetilde{\cD}_{ij}(.)$ satisfies the following equality for every MS $i$ and $t\geq 0$ $w.p.$ 1:\\
$\widetilde{X}_{i}(t)=\lambda_{i}t-\sum_{j\in\sN, j \neq i}\widetilde{\cD}_{ij}(t)-\widetilde{\cD}_{iD}(t)$.
\end{lem}
\begin{IEEEproof}Since $\widetilde{X}_{i}(.),  \widetilde{\cA}_{i}(.),  \widetilde{\cD}_{i0}(.) \text{ and } \widetilde{\cD}_{ij}(.)$ are fluid limits, there exists a sequence $r_{n_{k}}$ such that $\lim_{k\to\infty}r_{n_{k}}=\infty$ and they are obtained as a uniform limits of functions $X_{i}^{r_{n_{k}}}(.),  \cA_{i}^{r_{n_{k}}}(.),  \cD_{i0}^{r_{n_{k}}}(.), \cD_{ij}^{r_{n_{k}}}(.)$, respectively. Now, from (\ref{eq:X_cumu}) it follows that for every $r_{n_{k}}$ and $t\geq 0$
\begin{align}
X_{i}^{r_{n_{k}}}(t)=X_{i}^{r_{n_{k}}}(0)+\cA_i^{r_{n_{k}}}(t)-\sum_{j\in\sN}\cD_{ij}^{r_{n_{k}}}(t)-\cD_{i0}^{r_{n_{k}}}(t). \nonumber
\end{align}
The results follow from Lemma \ref{lem:lem1} after taking the limit $k\to \infty$ on both sides of the above equality.
\end{IEEEproof}
\begin{lem}
\label{lem:lem3}
Any fluid limit $\widetilde{Y}_{i}(.),  \widetilde{\cA}_{i}(.),  \widetilde{\cD}_{ji}(.), \widetilde{\cD}_{iR}(.)$ satisfies the following equality 	for every MS $i$ and $t\geq 0$ $w.p.$ 1:\\
$\widetilde{Y}_{i}(t)=\sum_{j\in\sN, j \neq i}\widetilde{\cD}_{ji}(t)-\widetilde{\cD}_{iR}(t)$.
\end{lem}
\begin{IEEEproof} The lemma can be proved in a similar way as Lemma~\ref{lem:lem2}. \end{IEEEproof} 
\begin{lem}
\label{lem:lem4}
Any fluid limit $\widetilde{U}_{i}(.), \widetilde{\cA}_{Ui}(.), \widetilde{\cD}_{Ui}(.)$ satisfies the following equality 	for every MS $i$ and $t\geq 0$ $w.p.$ 1:\\
$\widetilde{U}_{i}(t) = \widetilde{\cA}_{Ui}(t) - \widetilde{\cD}_{Ui}(t)$.
\end{lem}

\begin{IEEEproof} The lemma can be proved in a similar way as Lemma~\ref{lem:lem2}.\end{IEEEproof}
Note that $\widetilde{\cD}_{Ui}(t)\leq \oP_{i}t$, as shown below:
\begin{align}
\widetilde{\cD}_{Ui}(t) &= \lim_{k\to\infty}\frac{\cD_{Ui}(r_{n_{k}}t)}{r_{n_{k}}}\nonumber \\
&=\lim_{k\to\infty}\frac{\cD_{Ui}(r_{n_{k}}t)t}{r_{n_{k}}t}\nonumber \\
&\leq\oP_{i}t \hspace{3mm}(\text{By SLLN and $\cD_{Ui}(n) \leq n\oP_{i}$}). 
\end{align}

\section{Proof of Theorem 1}
The proof is done in two stages: (a)~$\occ_{IID} \subseteq \Lambda_{IID}$ and (b)~$\Lambda_{IID} \subseteq \occ_{IID}$ using the supplementary results from Appendix A. 
 
First we prove stage (a). Suppose, there exists $\bm{\lambda} \in \occ_{IID}$. By \cite[Definition 3]{ICC}, there exists a randomized policy $\Delta$ such that $\bm{\lambda}$ is rate stable under $\Delta$.
Fix any such $\Delta$ and let us define the following quantities:

\noindent
$\bullet$ Define indicator $\mI_{s}(n)=1$ if $s_{IID}(n)=s$, and 0 otherwise. 

\noindent
$\bullet$ Define indicator $\mI_{k}^{\Delta}(n)=1$ when $\bm{\l}^{\Delta}(n)=\bm{\l}_k$, and 0 otherwise. 

\noindent $\bullet$
$\mI_{i_{X}}^{\Delta}(n)=1$, when the packets from the own queue of MS $i$ is served on link $(i,0)$ in slot $n$, and 0 otherwise. 

Additionally,
\begin{align}
\label{eq:ind1}
&\lim_{T\to \infty}\frac{1}{T}\sum_{n=1}^{T}\mI_{s}(n)\mI_{k}^{\Delta}(n)=x_{sk}^{\Delta} \textit{ w.p. }1,  \\
\label{eq:ind2}
&\lim_{T\to \infty}\frac{1}{T}\sum_{n=1}^{T}\mI_{s}(n)\mI_{k}^{\Delta}(n)\mI_{i_{X}}^{\Delta}(n)=y_{i_{X}sk}^{\Delta} \textit{ w.p. }1, \\
\label{eq:ind3}
&\lim_{T\to \infty}\frac{1}{T}\sum_{n=1}^{T}\mI_{s}(n)\mI_{k}^{\Delta}(n)(1-\mI_{i_{X}}^{\Delta}(n))=y_{i_{Y}sk}^{\Delta} \textit{ w.p. }1,  \\
\label{eq:ind4}
&\oP_{i}^{\Delta}=\lim_{T \to \infty}\frac{1}{T}\sum_{n=1}^{T}\sum_{j \in [N], j \neq i}\l_{ij}^{\Delta}(n). 
\end{align}
Assume that the system is ergodic such that the above limits exist. Equation~(\ref{eq:ind1}) refers to fraction of time network topology of system is $s$ and power vector $\bl_{k}$ has been selected by policy $\Delta$. Equation~(\ref{eq:ind2}) refers to fraction of time  own queue of MS $i$ has been selected when topology is $s$ and power vector $\bl_{k}$ is selected by policy $\Delta$. Equation~(\ref{eq:ind3}) refers to similar fraction for relay queue of MS $i$. Equation~(\ref{eq:ind4}) refers to power consumed at MS $i$ under policy $\Delta$.

Let  $w_{sk}^{\prime\Delta}$ be the fraction of time power vector $\bl_{k}$ has been selected under policy $\Delta$ given that network topology is $s$. Thus, 
\begin{align}
w_{sk}^{\prime\Delta}  = \frac{x_{sk}^{\Delta}}{\pi_{s}} = \lim_{T\to \infty}\frac{\frac{1}{T}\sum_{n=1}^{T}\mI_{s}(n)\mI_{k}^{\Delta}(n)}{\frac{1}{T}\sum_{n=1}^{T}\mI_{s}(n)}. 
\end{align}
Similarly, let $q_{isk}^{\prime}$ be  the fraction of time own queue has been selected for transmission given that power vector is $\bl_{k}$ and topology is at $s$.
\begin{align}
q_{isk}^{\prime} = \frac{y_{i_{X}sk}^{\Delta}}{x_{sk}^{\Delta}} = \frac{\lim_{T\to \infty}\frac{1}{T}\sum_{n=1}^{T}\mI_{s}(n)\mI_{k}^{\Delta}(n)\mI_{i_{X}}^{\Delta}(n)}{\lim_{T\to \infty}\frac{1}{T}\sum_{n=1}^{T}\mI_{s}(n)\mI_{k}^{\Delta}(n)}.  
\end{align}   
\remove{\begin{align}
\frac{\lim_{T\to \infty}\frac{1}{T}\sum_{n=1}^{T}\mI_{s}(n)\mI_{k}^{\Delta}(n)(1-\mI_{i_{X}}^{\Delta}(n))}{\lim_{T\to \infty}\frac{1}{T}\sum_{n=1}^{T}\mI_{s}(n)\mI_{k}^{\Delta}(n)}= \frac{y_{i_{Y}sk}^{\Delta}}{x_{sk}^{\Delta}} =  \nonumber
\end{align}}  
Then, $1-q_{isk}^{\prime}$  the fraction of time packets from relay queue has been selected for transmission under same condition.

Recall that we assumed that $\bm{\lambda} \in \occ_{IID}$. So, from \cite[Definition 2]{ICC} and \cite[Definition 3]{ICC}, the policy $\Delta$ is feasible, i.e., $\oP_{i}^{\Delta}\leq \oP_{i}$. Now,
\begin{align}
\oP_{i}^{\Delta}&=\lim_{T \to \infty}\frac{1}{T}\sum_{n=1}^{T}\sum_{j \in [N], j \neq i}\l_{ij}^{\Delta}(n) \nonumber \\
&=\lim_{T \to \infty}\frac{1}{T}\sum_{n=1}^{T}\sum_{\bs\in \cS}\sum_{k=1}^{M}\sum_{j\in[N], j \neq i}\l_{ijk}\mI_{s}(n)\mI_{k}^{\Delta}(n) \nonumber \\
&=\sum_{\bs\in \cS}\sum_{k=1}^{M}\sum_{j\in[N], j \neq i}\l_{ijk}\Big(\lim_{T\to \infty}\frac{1}{T}\sum_{n=1}^{T}\mI_{s}(n)\mI_{k}^{\Delta}(n)\Big) \nonumber \\
&=\sum_{\bs\in \cS}\sum_{k=1}^{M}\sum_{j\in[N], j \neq i}\l_{ijk}x_{sk}^{\Delta} \nonumber \\
&=\sum_{\bs\in \cS}\sum_{k=1}^{M}\pi_{s}w_{sk}^{\prime\Delta}\sum_{j\in[N], j \neq i}\l_{ijk} \text{ w.p. } 1.\nonumber
\end{align}
Then, $\sum_{\bs\in \cS}\sum_{k=1}^{M}\pi_{s}w_{sk}^{\prime\Delta}\sum_{j\in[N], j \neq i}\l_{ijk}\leq \oP_{i}$. Similarly, we have
\begin{align}
\omu_{ij}^{\Delta}&=\lim_{T\to \infty}\frac{1}{T}\sum_{n=1}^{T}\mu_{ij}^{\Delta}(n) \nonumber  \\
\remove{&=\lim_{T\to\infty}\frac{1}{T}\sum_{n=1}^{T}\sum_{\bs\in \cS}\sum_{k=1}^{M}\mu_{ij}(s,\l_{k})\mI_{s}(n)\mI_{k}^{\Delta}(n) \nonumber\\ \nonumber
&=\sum_{\bs\in \cS}\sum_{k=1}^{M}\mu_{ij}(s,\l_{k})x_{sk}^{\Delta}\\ \nonumber}
&=\sum_{\bs\in \cS}\sum_{k=1}^{M}\mu_{ij}(s,\l_{k})\pi_{s}w_{sk}^{\prime\Delta} w.p. 1, \nonumber
\end{align}
\begin{align}
\overline{\mu}_{iR}^{\Delta}&=\lim\inf_{\10i T\to\infty}\frac{1}{T} \sum_{t=1}^T \mu_{i0}^{\Delta}(n) [1-\mathbb{I}_i^{\Delta}(n)] \nonumber \\
\remove{
&=\lim_{T\to\infty}\frac{1}{T}\sum_{n=1}^{T}\sum_{\bs\in \cS}\sum_{k=1}^{M}\mu_{i0}(s,\l_{k})\mI_{s}(n)\mI_{k}^{\Delta}(n)(1-\mI_{i_{X}}^{\Delta}(n)) \nonumber \\
&=\sum_{\bs\in \cS}\sum_{k=1}^{M}\mu_{i0}(s,\l_{k})y_{i_{Y}sk} \nonumber \\
&=\sum_{\bs\in \cS}\sum_{k=1}^{M}\mu_{i0}(s,\l_{k})(1-q_{isk}^{\prime})x_{sk}^{\Delta} \nonumber \\}
&=\sum_{\bs\in \cS}\sum_{k=1}^{M}\mu_{i0}(s,\l_{k})(1-q_{isk}^{\prime})\pi_{s}w_{sk}^{\prime\Delta} w.p. 1.\nonumber
\end{align}
 Recall that from  \cite[Definition 2]{ICC}, $\sum_{j\in [N], j \neq i}\omu_{ji}^{\Delta}\leq \overline{\mu}_{iR}^\Delta $. Then,
\begin{align}
\sum_{j \in [N], j \neq i}\sum_{\bs\in \cS}&\sum_{k=1}^{M}\mu_{ji}(s,\l_{k})\pi_{s}w_{sk}^{\prime\Delta}\leq \nonumber \\
&\sum_{\bs\in \cS}\sum_{k=1}^{M}\mu_{i0}(s,\l_{k})(1-q_{isk}^{\prime})\pi_{s}w_{sk}^{\prime\Delta}. \nonumber
\end{align}
Also, from \cite[Definition 2]{ICC},  $\lambda_{i}\leq \overline{\mu}_{i0}^\Delta+\sum_{j \in [N], j \neq i}\omu_{ij}^{\Delta}$. Doing the analysis in a similar  way, we get \\
\begin{align}
\sum_{s}\sum_{k}\pi_{s}w_{sk}^{\prime\Delta} [ \mu_{i0}(s,\l_{k})q_{isk}^{\prime}+\sum_{j\in [N], j \neq i}\mu_{ij}(s,\l_{k})]\geq \lambda_{i}. \nonumber
\end{align} 
Hence, we have shown that there exists  $w_{sk}^{\prime\Delta}$ and $q_{isk}^{\prime}$ which satisfy the constraints for $\Lambda_{IID}$ and thus $\bm{\lambda} \in \Lambda_{IID}$. So, $\occ_{IID} \subseteq \Lambda_{IID}$.

Now, we have to prove $\Lambda_{IID}\subseteq\occ_{IID}$. For any $\bm{\lambda}\in\Lambda_{IID}$, we have a randomized resource allocation policy with appropriate $w_{sk}$ and $q_{isk}$, $\forall s \in \cS, \bl_{k} \in \cL$ and $i \in [N]$ \remove{which selects a power vector $\bl_{k}$ given a topology $s$ with probability $w_{sk}$ and chooses own queue at a MS $i$ when power vector is $k$ in topology $s$  with probability $q_{isk}$ } satisfying \cite[Equation 9, Equation 10, Equation 11, Equation 12, Equation 13]{ICC}. We will try to show that this randomized policy is rate stable. Assume that the fluid limits $\widetilde{X}_{i}(0)= \widetilde{Y}_{i}(0)=\widetilde{U}_{i}(0)=0$. Then, we will try to show that $\widetilde{X}_{i}(t)= \widetilde{Y}_{i}(t)=\widetilde{U}_{i}(t)=0$ for every $t>0$. Towards this end, in order to measure the aggregate network congestion, we define a Lyapunov function $V(t)$ as a sum of the squares of the fluid limits of the individual queue lengths:
\begin{align}
\label{eq:Lya}
&V(t)=\sum_{i}\widetilde{X}_{i}^{2}(t)+\sum_{i}\widetilde{Y}_{i}^{2}(t)+\sum_{i}\widetilde{U}_{i}^{2}(t). 
\end{align}
From our assumption, $V(0)=0$. Our task will be to show $V(t)=0, \forall t>0$. To achieve that, we just need to show that $V'(t)\leq 0, \forall t>0$ after which we appeal to the \cite[Lemma 1]{DP}. This lemma is provided here for reference.
\begin{lem}
\label{lem:lemma5}
Let $f:[0,\infty) \rightarrow [0,\infty)$ be an absolutely continuous function with $f(0)=0$. Assume that whenever  $f(t)>0$ and f is differentiable at $t$, $f'(t)\leq 0$ for almost every $t$ (wrt Lebesgue measure). Then $f(t)=0$ for almost every $t\geq 0$.
\end{lem}
The fact that fluid limits are Lipschitz continuous guarantees $V(t)$ to be differentiable almost everywhere. Applying Leibniz's rule for differentiation to (\ref{eq:Lya}) and taking only right-hand limit, we have
\begin{align}
\label{eq:Lyadif}
V'(t)&=\sum_{i}\widetilde{X}_{i}(t)\lim_{\epsilon \to 0^{+}}\frac{\widetilde{X}_{i}(t+\epsilon)-\widetilde{X}_{i}(t)}{\epsilon} \nonumber \\
&+\sum_{i}\widetilde{Y}_{i}(t)\lim_{\epsilon \to 0^{+}}\frac{\widetilde{Y}_{i}(t+\epsilon)-\widetilde{Y}_{i}(t)}{\epsilon} \nonumber \\
&+\sum_{i}\widetilde{U}_{i}(t)\lim_{\epsilon \to 0^{+}}\frac{\widetilde{U}_{i}(t+\epsilon)-\widetilde{U}_{i}(t)}{\epsilon}.
\end{align}
Suppose $V(0)=0$. We will try to prove that $V(t)=0$ for some $t>0$. Assume that $V(t)>0$ for some $t>0$. Now, we have the following cases:\\
\begin{enumerate}
\item Only one of the three queues is non-zero.
\item Any two of  the three queues is non-zero.
\item All of the three queues are non-zero.
\end{enumerate}
\subsubsection{Case 1} \hspace{2mm}Assume that $\widetilde{X}_{i}(t)>0$ for some $t>0$. Without loss of generality, we take 
\begin{align}
\label{eq:X1}
\lim_{\epsilon \to 0^{+}}&\frac{\widetilde{X}_{i}(t+\epsilon)-\widetilde{X}_{i}(t)}{\epsilon} > 0,
\end{align}
for some $\epsilon>0$. Let $a=\min_{t' \in [t,t+\epsilon]}\widetilde{X}_{i}(t') > 0$. Thus, for large enough $k$, we have $\frac{X_{i}(r_{n_{k}}t')}{r_{n_{k}}}\geq a \hspace{2mm} \forall t' \in [t,t+\epsilon] \text{ and } ar_{n_{k}} > \mu_{max} $. Then, $X_{i}(r_{n_{k}}t') \geq \mu_{max} \hspace{2mm} \forall t' \in [t,t+\epsilon]$. Since we are guaranteed that minimum queue length in this interval is $\mu_{max}$, so we can write
\begin{align}
&\lim_{\epsilon \to 0^{+}}\frac{\widetilde{X}_{i}(t+\epsilon)-\widetilde{X}_{i}(t)}{\epsilon}\nonumber \\
&=\lim_{\epsilon \to 0^{+}}\frac{\widetilde{\cA}_{i}(\epsilon)-\sum_{j\in\sN}\widetilde{\cD}_{ij}(\epsilon)-\widetilde{\cD}_{i0}(\epsilon)}{\epsilon} \nonumber \\
&=\lambda_{i}- \nonumber \\
&\left[\sum_{s\in\cS}\sum_{k=1}^M\pi_{s}w_{sk} \left[ \mu_{i0}(s,\l_{k})q_{isk}+\sum_{j\in [N], j \neq i}\mu_{ij}(s,\l_{k})\right]\right]. \nonumber 
\end{align}  
Then, by (\ref{eq:X1})
\begin{align}
& \lambda_{i}\geq \nonumber \\  &\left[\sum_{s\in\cS}\sum_{k=1}^M\pi_{s}w_{sk} \left[ \mu_{i0}(s,\l_{k})q_{isk}+\sum_{j\in [N], j \neq i}\mu_{ij}(s,\l_{k})\right]\right], \nonumber 
\end{align}
which is a contradiction as we have assumed that $\bm{\lambda} \in \Lambda_{IID}$. Similarly, we can prove that $\widetilde{Y}_{i}(t')$ is decreasing for every $t'$. 

Suppose
\begin{align}
\label{ref:contra2}
&\lim_{\epsilon \rightarrow 0^{+}}\frac{\widetilde{U}_{i}(t+\epsilon)-\widetilde{U}_{i}(t)}{\epsilon}>0.
\end{align}
Same as before, for large enough $k$, we can have $U_{i}(r^{n_{k}}t') \geq \oP_{i}\hspace{2mm} \forall t' \in [t,t+\epsilon]$. So, we can always expect departure from token queue to take place at maximum  capacity every instant in this interval. Then, 
\begin{align}
&\lim_{\epsilon \rightarrow 0^{+}}\frac{\widetilde{U}_{i}(t+\epsilon)-\widetilde{U}_{i}(t)}{\epsilon}\nonumber \\
&=-\oP_{i}+\sum_{s\in\cS}\sum_{k=1}^M\pi_{s}w_{sk}\sum_{j\in[N], j \neq i}\l_{ijk}. \nonumber   
\end{align}
Then, by (\ref{ref:contra2}), we have 
\begin{align}
\sum_{s\in\cS}\sum_{k=1}^M\pi_{s}w_{sk}\sum_{j\in[N], j \neq i}\l_{ijk} \geq \oP_{i}, \nonumber
\end{align}
which is a contradiction as we assumed $\bm{\lambda} \in \Lambda_{IID}$. Thus, $V'(t)\leq 0 \hspace{2mm}\forall t>0$.

\subsubsection{Case 2}Again, w.l.o.g, assume that queues $\widetilde{X}_{i}(t)$ and $\widetilde{Y}_{i}(t)$ are non-zero for some $t>0$. From case 1, we have 
\begin{align}
\lim_{\epsilon \to 0^{+}}&\frac{\widetilde{X}_{i}(t+\epsilon)-\widetilde{X}_{i}(t)}{\epsilon} \leq 0,\nonumber \\
\lim_{\epsilon \to 0^{+}}&\frac{\widetilde{Y}_{i}(t+\epsilon)-\widetilde{Y}_{i}(t)}{\epsilon} \leq 0.\nonumber
\end{align}
So, $V'(t)\leq 0 \hspace{2mm}\forall t>0$. Similarly, we can take other possible combinations of queues, two at a time, and we can show the same result.

\subsubsection{Case 3}Again, without loss of generality, assume that queues $\widetilde{X}_{i}(t)$, $\widetilde{Y}_{i}(t)$ and $\widetilde{U}_{i}(t)$ are non-zero for some $t>0$. From case 1, we have 
\begin{align}
\lim_{\epsilon \to 0^{+}}&\frac{\widetilde{X}_{i}(t+\epsilon)-\widetilde{X}_{i}(t)}{\epsilon} \leq 0,\nonumber \\
\lim_{\epsilon \to 0^{+}}&\frac{\widetilde{Y}_{i}(t+\epsilon)-\widetilde{Y}_{i}(t)}{\epsilon} \leq 0,\nonumber \\
\lim_{\epsilon \to 0^{+}}&\frac{\widetilde{U}_{i}(t+\epsilon)-\widetilde{U}_{i}(t)}{\epsilon} \leq 0.\nonumber
\end{align}
Thus, $V'(t)\leq 0 \hspace{2mm}\forall t>0$ for every case. 

Hence, by Lemma~\ref{lem:lemma5}, $V(t)=0$ for $t\geq 0$. However, $\widetilde{X}_{i}(t)$, $\widetilde{Y}_{i}(t)$ and $\widetilde{U}_{i}(t)$ are non-negative reals. Therefore, $\widetilde{X}_{i}(t)=0$, $\widetilde{Y}_{i}(t)=0$ and $\widetilde{U}_{i}(t)=0 \hspace{2mm} \forall i \in [1,..,N]$. Then using Lemma~\ref{lem:lem1}, \ref{lem:lem2}, \ref{lem:lem3} and \ref{lem:lem4}, we get 
\begin{align}
&\widetilde{\cA}_{i}(t)=\sum_{j\in[N], j \neq i}\widetilde{\cD}_{ij}(t)+\widetilde{\cD}_{i0}(t), \nonumber\\
&\sum_{j\in[N], j \neq i}\widetilde{\cD}_{ji}(t)=\widetilde{\cD}_{iR}(t),\nonumber \\
\label{eq:power1}
&\widetilde{\cA}_{Ui}(t) = \widetilde{\cD}_{Ui}(t)\leq \oP_{i}t. 
\end{align}
Writing the expression for the fluid limits and making appropriate manipulations, we get
\begin{align}
&\lim_{k\to\infty}\frac{\sum_{j \in [N], j \neq i}\cD_{ij}(r_{n_{k}}t)+\cD_{i0}(r_{n_{k}}t)}{r_{n_{k}}t}=\lambda_{i}, \nonumber\\
&\lim_{k\to\infty}\frac{\sum_{j \in [N], j \neq i}\cD_{ji}(r_{n_{k}}t)}{r_{n_{k}}t}=\lim_{k\to\infty}\frac{\cD_{iR}(r_{n_{k}}t)}{r_{n_{k}}t}. \nonumber
\end{align}
We have \textit{Average Departure rate = Average Arrival rate} for every queue. But, its a fact that \textit{Average Departure rate $\leq$ Average Service rate} for any queue. Let $\omu_{ij}^{\Delta},\omu_{i0}^{\Delta}$ and $\omu_{iR}^{\Delta}$ be the average service rate of link $(i,j)$, and of link $(i,0)$ for \textit{own queue} and \textit{relay queue} of MS $i$, respectively as defined in \cite[Equation 6, Equation 7, Equation 8]{ICC} under the randomised resource allocation policy $\Delta$. Then, by strong law of large numbers and assuming large queue backlog, we have
\begin{align}
&\sum_{j\in [N], j \neq i}\omu_{ij}^{\Delta}+\omu_{i0}^{\Delta} \geq \lambda_{i}, \nonumber \\
&\sum_{j\in [N], j \neq i}\omu_{ji}^{\Delta} \leq \omu_{iR}^{\Delta} \nonumber.
\end{align}
Thus, conditions $(a)$ and $(b)$ of \cite[Definition 2]{ICC} are met. Furthermore, (\ref{eq:power1}) can be written as:
\begin{align}
\frac{\widetilde{A}_{Ui}(t)}{t} = \oP_{i}^{\Delta} \leq \oP_{i} \hspace{2mm}(\text{by SLLN}). \nonumber
\end{align}
This implies the randomized policy is feasible. Hence, for any $\bm{\lambda}\in\Lambda_{IID}$, we have shown the existence of a randomized policy which is rate stable for that $\bm{\lambda}$. Then using \cite[Definition 3]{ICC}, $\bm{\lambda}\in\occ_{IID}$ and hence, $\Lambda_{IID}\subseteq\occ_{IID}$. Thus, $\Lambda_{IID}=\occ_{IID}$.

\section{Proof of Lemma 1}
\label{sec:ProofOptIID}
We assume that $\bm{\lambda}$ is stabilizable. Note that for policy $\Delta_{IID}^{o}$, $\{\textbf{X}(n),\textbf{Y}(n),\textbf{U}(n)\}_{n\geq 1}$ is an irreducible, aperiodic and countable Markov chain. Now, define the following Lyapunov function: 
\begin{align}
\label{eq:Lyamod}
    f(n) = \sum_{i=1 \in [N]}\bigg[(X_{i}(n))^{2} + (Y_{i}(n))^{2} + (U_{i}(n))^{2}\bigg].
\end{align} 
Recall that $\mu_{max}$ be the maximum possible transmission rate across any link in the network at any topology and $P_{max}$ be the maximum transmit power for any MS in the network. Let $M \myeq NA_{max}^{2} + N(N\mu_{max})^{2} + \sum_{i \in [N]} \oP_{i} + NP_{max}^{2}$. It follows that
\begin{align}
    & f(n+1) - f(n)  \nonumber \\
    & \leq M - 2\sum_{i \in [N]}U_{i}(n)\oP_{i} + 2\sum_{i \in [N]}X_{i}(n)A_{i}(n) \nonumber \\
    &-2\sum_{i \in [N]}\Bigg[\sum_{\substack{j \in [N]\\ j\neq i}}\bigg( X_{i}(n) - Y_{i}(n) \bigg)\mu_{ij}(s_{IID}(n),\bl^{\Delta_{IID}^{o}}(n)) \nonumber \\
    & \hspace{1cm} + \max(X_{i}(n),Y_{i}(n))\mu_{i0}(s_{IID}(n),\bl^{\Delta_{IID}^{o}}(n))  \nonumber \\
    & \hspace{1cm} - U_{i}(n)\sum_{\substack{j=0 \\ j\neq i}}^{N}\l_{ij}^{\Delta_{IID}^{o}}(n)\Bigg]. \nonumber
\end{align}

Thus,
\begin{align}
   &\mathop{\mathbb{E}}[f(n+1)-f(n)|\mathbf{X}(n),\mathbf{Y}(n),\mathbf{U}(n)] \nonumber \\
   & \leq M - 2\sum_{i \in [N]}U_{i}(n)\oP_{i} + 2\sum_{i \in [N]}X_{i}(n)\lambda_{i}\nonumber \\
   & - 2\sum_{i \in [N]}\mathop{\mathbb{E}}\Bigg[\sum_{\substack{j \in [N]\\ j\neq i}}\bigg( X_{i}(n) - Y_{i}(n) \bigg)\mu_{ij}(s_{IID}(n),\bl^{\Delta_{IID}^{o}}(n)) \nonumber \\
   &  + \max\{X_{i}(n),Y_{i}(n)\}\mu_{i0}(s_{IID}(n),\bl^{\Delta_{IID}^{o}}(n))  \nonumber \\ 
   & - U_{i}(n)\sum_{j=0,j\neq i}^{N}\l_{ij}^{\Delta_{IID}^{o}}(n) \mid \mathbf{X}(n),\mathbf{Y}(n),\mathbf{U}(n) \Bigg] \nonumber \\
   & = M - 2\sum_{i \in [N]}U_{i}(n)\oP_{i} + 2\sum_{i \in [N]}X_{i}(n)\lambda_{i} \nonumber \\
   & \hspace{0.7cm} - 2\mathop{\mathbb{E}}[W(\textbf{X}(n),\textbf{Y}(n),\textbf{U}(n),s_{IID}(n),\bl^{\Delta_{IID}^{o}}(n))\mid \nonumber \\
   \label{eq:ProofOptIID1}
   & \hspace{5.5cm} \mathbf{X}(n),\mathbf{Y}(n),\mathbf{U}(n)],
\end{align}
where,
\begin{align}
    &W(\textbf{X}(n),\textbf{Y}(n),\textbf{U}(n),s_{IID}(n),\bl^{\Delta_{IID}^{o}}(n)) \nonumber \\
    &=\sum_{i \in [N]}\Bigg[\sum_{\substack{j \in [N]\\j\neq i}}\bigg( X_{i}(n) - Y_{i}(n) \bigg)\mu_{ij}(s_{IID}(n),\bl^{\Delta_{IID}^{o}}(n)) \nonumber \\
    & + \max\{X_{i}(n),Y_{i}(n)\}\mu_{i0}(s_{IID}(n),\bl^{\Delta_{IID}^{o}}) \nonumber \\
    & - U_{i}(n)\sum_{\substack{j=0\\j\neq i}}^{N}\l_{ij}^{\Delta_{IID}^{o}}(n)\Bigg].\nonumber
\end{align}

Now, since $\bm{\lambda}$ is stabilizable, we can obtain a randomized policy $\Delta_{IID}$ satisfying \cite[Equation (9)-(13)]{ICC}. From definition of $\Delta_{IID}^{o}$ (\cite[Equation 14]{ICC}), for every $\Delta_{IID}$ and $n \geq 0$
\begin{align}
\label{eq:inequality1}
    &\mathbb{\mathop{E}}[W(\textbf{X}(n),\textbf{Y}(n),\textbf{U}(n),s_{IID}(n),\bl^{\Delta_{IID}^{o}}(n)) \nonumber \\
    & \hspace{4.5cm}\mid \mathbf{X}(n),\mathbf{Y}(n),\mathbf{U}(n)] \nonumber \\
    &  \hspace{0.6cm}\geq \mathbb{\mathop{E}}[W(\textbf{X}(n),\textbf{Y}(n),\textbf{U}(n),s_{IID}(n),\bl^{\Delta_{IID}}(n)) \nonumber \\
    & \hspace{4.5cm}\mid \mathbf{X}(n),\mathbf{Y}(n),\mathbf{U}(n)].
\end{align}
Thus, from (\ref{eq:ProofOptIID1}), 
\begin{align}
   &\mathop{\mathbb{E}}[f(n+1)-f(n)|\mathbf{X}(n),\mathbf{Y}(n),\mathbf{U}(n)] \nonumber \\  
   & \leq M - 2\sum_{i \in [N]}U_{i}(n)\oP_{i} + 2\sum_{i \in [N]}X_{i}(n)\lambda_{i} \nonumber \\
   & \hspace{0.7cm} - 2\mathop{\mathbb{E}}[W(\textbf{X}(n),\textbf{Y}(n),\textbf{U}(n),s_{IID}(n),\bl^{\Delta_{IID}}(n))\mid \nonumber \\
   \label{eq:ProofOptIID2}
   & \hspace{5.5cm} \mathbf{X}(n),\mathbf{Y}(n),\mathbf{U}(n)].   
\end{align}
there exists some $\delta > 0$ such that in \cite[Equation (11)-(13)]{ICC}, we can write
\begin{align}
\label{eq:DeltaEqn1}
& \sum_{s\in\cS}\sum_{k=1}^M\pi_{s}w_{sk} \left[ \mu_{i0}(s,\l_{k})q_{isk}+\sum_{j\in [N], j \neq i}\mu_{ij}(s,\l_{k})\right] \nonumber \\
& \hspace{2cm}\geq  \lambda_{i} + \delta, \\
&\sum_{s\in\cS}\sum_{k=1}^M\pi_{s}(1-q_{isk})w_{sk}\mu_{i0}(s,\l_{k}) \quad \nonumber \\
\label{eq:DeltaEqn2}
& \qquad \hspace{1cm}\geq \sum_{s\in\cS}\sum_{k=1}^M\sum_{j\in [N], j \neq i}\pi_{s}w_{sk}\mu_{ji}(s,\l_{k}) + \delta, \\
\label{eq:DeltaEqn3}
& \sum_{s\in\cS}\sum_{k=1}^M\pi_{s}w_{sk}\sum_{j\in[N], j \neq i}\l_{ijk} + \delta \leq \oP_{i}.  
\end{align}
Thus, we have
\begin{align}
    &\mathop{\mathbb{E}}[W(\textbf{X}(n),\textbf{Y}(n),\textbf{U}(n),s_{IID}(n),\bl^{\Delta_{IID}}(n))\mid \nonumber \\
    &\hspace{5.5cm} \mathbf{X}(n),\mathbf{Y}(n),\mathbf{U}(n)]  \nonumber \\
    & \geq \sum_{i \in [N]}\Bigg[ X_{i}(n)(\lambda_{i} + \delta) + Y_{i}(n)\delta - U_{i}(n)(\oP_{i} - \delta) \Bigg]. \nonumber 
\end{align}
Hence, from (\ref{eq:ProofOptIID2})
\begin{align}
    &\mathop{\mathbb{E}}[f(n+1)-f(n)|\mathbf{X}(n),\mathbf{Y}(n),\mathbf{U}(n)] \nonumber \\
    & \leq M - 2\delta\sum_{i \in [N]}(X_{i}(n) + Y_{i}(n) + U_{i}(n)).
\end{align}
Let $\mathcal{A} = \{\{\mathbf{X}(n),\mathbf{Y}(n),\mathbf{U}(n)\}:\sum_{i \in [N]}( X_{i}(n) + Y_{i}(n) +  U_{i}(n) ) \leq (M+1)/2\delta \}$. Then,
\begin{align}
    \mathop{\mathbb{E}}& [f(n+1)-f(n)|\mathbf{X}(n),\mathbf{Y}(n),\mathbf{U}(n)] \nonumber \\
    &< \begin{cases}
        \infty, & \forall \text{ } \{\mathbf{X}(n),\mathbf{Y}(n),\mathbf{U}(n)\} \\
        -1, & \text{if } \{\mathbf{X}(n),\mathbf{Y}(n),\mathbf{U}(n)\} \notin \mathcal{A}. 
  \end{cases}
\end{align}
Thus, since $\mathopen|\mathcal{A}\mathclose|$ is finite, by Foster's Theorem \cite[Theorem~2.2.3]{GVM}, $\{\mathbf{X}(n), \mathbf{Y}(n), \mathbf{U}(n)\}_{n \geq 0}$ is positive recurrent, and for each queue the expected queue length under its stationary distribution is finite. Thus, the system is stable under $\Delta_{IID}^{o}$.

\section{Proof of Lemma 2}
\label{sec:ProofTstepstable}
Let $\bm{\lambda}$ be stabilizable and 
\begin{align}
\label{eq:Lyamod1}
    f(n) = \sum_{i=1 \in [N]}\bigg[(X_{i}(n))^{2} + (Y_{i}(n))^{2} + (U_{i}(n))^{2}\bigg].
\end{align}
Let $M_{1}\myeq N[(TA_{max})^2 + (T\mu_{max})^2 + (N^2 + 1)T^2\mu_{max}^2 + 2T^2 \overline{P}_{i}^2]$. Using the analysis similar to the one used for obtaining (\ref{eq:ProofOptIID1})
\begin{align}
        & \mathop{\mathbb{E}}[f((K+1)T)-f(KT)|\mathbf{X}(KT),\mathbf{Y}(KT),\mathbf{U}(KT)] \nonumber  \\
        & \leq M_{1} - \nonumber \\
        &  2\sum_{n = KT}^{\substack{(K+1)T 
        \\- 1}}\mathop{\mathbb{E}}\bigg[\sum_{i,j}(X_{i}(KT)-Y_{j}(KT)\mu_{ij}(s_{IID}(n),\bl^{\Delta_{IID}^{T}}(n)) \nonumber \\ 
        & + \sum_{i \in [N]}\max{(X_{i}(KT),Y_{i}(KT))}\mu_{i0}(s_{IID}(n),\bl^{\Delta_{IID}^{T}}(n)) - \nonumber \\
        &  \sum_{i \in [N]}\sum_{\substack{j=0\\j\neq i}}^{N}U_{i}(KT)\l_{ij}^{\Delta_{IID}^{T}}(n)|\mathbf{X}(KT),\mathbf{Y}(KT),\mathbf{U}(KT)\bigg] \nonumber  \\
        & + 2T\sum_{i \in [N]}\lambda_{i}X_{i}(KT) -2T\sum_{i \in [N]}\overline{P}_{i}U_{i}(KT) \nonumber \\
        & =  M_{1} - 2\times \nonumber \\ 
        & \sum_{n = KT}^{\substack{(K+1)T \\- 1}}\mathop{\mathbb{E}}\bigg[W(\mathbf{X}(KT),\mathbf{Y}(KT),\mathbf{U}(KT),s_{IID}(n),\bl^{\Delta_{IID}^{T}}(n)) \nonumber \\
        &\hspace{4cm}\mid \mathbf{X}(KT),\mathbf{Y}(KT),\mathbf{U}(KT)\Bigg] \nonumber \\
        \label{eq:inequality3}
        & + 2T\sum_{i \in [N]}\lambda_{i}X_{i}(KT) - 2T\sum_{i \in [N]}\overline{P}_{i}U_{i}(KT),
\end{align}
where, 
\begin{align}
    &W(\mathbf{X}(KT),\mathbf{Y}(KT),\mathbf{U}(KT),s_{IID}(n),\bl^{\Delta_{IID}^{T}}(n))  = \nonumber \\
    & \sum_{i \in [N]}\Bigg[\sum_{\substack{j \in [N]\\j\neq i}}\bigg( X_{i}(KT) - Y_{i}(KT) \bigg)\mu_{ij}(s_{IID}(n),\bl^{\Delta_{IID}^{T}}(n)) \nonumber \\
    &  + \max(X_{i}(KT),Y_{i}(KT))\mu_{i0}(s_{IID}(n),\bl^{\Delta_{IID}^{T}}(n)) \nonumber \\
    &  - U_{i}(KT)\sum_{j=0,j\neq i}^{N}\l_{ij}^{\Delta_{IID}^{T}}(n)\Bigg]. \nonumber 
\end{align}
Our next step is to find a relationship between $W(\mathbf{X}(KT),\mathbf{Y}(KT),\mathbf{U}(KT),s_{IID}(n),\bl^{\Delta_{IID}^{T}}(n))$ and $W(\mathbf{X}(n),\mathbf{Y}(n),\mathbf{U}(n),s_{IID}(n),\bl^{\Delta_{IID}^{o}}(n))$. Towards that end, we have the following inequalities 
\begin{align}
\label{eq:Xrel1}
    &X_{i}(n) \leq  X_{i}(KT) + \sum_{t = KT}^{n-1}A_{i}(t) \leq X_{i}(KT) + TA_{max}, \\
    &X_{i}(n) \geq  X_{i}(KT) - \sum_{t=KT}^{n-1}\sum_{j \in [N]}\mu_{ij}(s(t),\bl(t) \nonumber \\
    & \hspace{2cm} - \sum_{t =KT}^{n-1}\mu_{i0}(s(t),\bl(t)) \nonumber \\ 
    \label{eq:Xrel2}
    & \hspace{1cm} \geq  X_{i}(KT) - T\mu_{max}, \\
\label{eq:Yrel3}
    &Y_{i}(n) \leq  Y_{i}(KT) + T(N-1)\mu_{max}, \\
\label{eq:Yrel4}
    &Y_{i}(n) \geq  Y_{i}(KT) - T\mu_{max},  \\
    &\max{(X_{i}(n),Y_{i}(n))}\mu_{i0}(s^{IID}(n),\bl(n)) \nonumber \\ 
    & \hspace{1cm}\leq  \max{(X_{i}(KT),Y_{i}(KT))}\mu_{i0}(s^{IID}(n),\bl(n)) \nonumber \\
\label{eq:max1}    
    & \hspace{2cm} + T\mu_{max}[A_{max} + N\mu_{max}],    \\
\label{eq:max2}
    &\max{(X_{i}(n),Y_{i}(n))}\mu_{i0}(s^{IID}(n),\bl(n)) \nonumber \\ 
    & \hspace{0.2cm} \geq  \max{(X_{i}(KT),Y_{i}(KT))}\mu_{i0}(s^{IID}(n),\bl(n)) - 2T\mu_{max}^2,    \\
\label{eq:Urel1}
    &U_{i}(n) \leq U_{i}(KT) + T\overline{P}_{i},\\
\label{eq:Urel2}
    &U_{i}(n) \geq U_{i}(KT) - T\overline{P}_{i},   
\end{align}
where we have used the identities $\max{(a+b,c)} \leq \max{(a,c)}+b$ and $\max{(a-b,c)}\geq\max{(a,c)}-b$ for obtaining (\ref{eq:max1}) and (\ref{eq:max2}). 

With $\bm{\lambda}$ being stabilizable, \cite[Lemma 1]{ICC} guarantees that policy $\bm{\lambda} \in \occ_{IID}^{\Delta_{IID}^{o}}$.  Let $\frac{M_{2}}{2}\myeq TN^2(A_{max}+\mu_{max})\mu_{max} + NT\mu_{max}[A_{max}+N\mu_{max}] + T\sum_{i \in [N]}\overline{P}_{i}^{2}$ From definition of $\Delta_{IID}^{T}$ (\cite[Equation 16]{ICC}) and from (\ref{eq:Xrel1})-(\ref{eq:Urel2}), for every $n \in [KT, (K+1)T-1]$
\begin{align}
\label{eq:Wcomp1}
        &W(\mathbf{X}(n),\mathbf{Y}(n),\mathbf{U}(n),s_{IID}(n),\bl^{\Delta_{IID}^{o}}(n))\nonumber  \\
        &\leq W(\mathbf{X}(KT),\mathbf{Y}(KT),\mathbf{U}(KT),s_{IID}(n),\bl^{\Delta_{IID}^{T}(n)}) + \frac{M_{2}}{2}.
\end{align}

From (\ref{eq:inequality3}) and (\ref{eq:Wcomp1}), and using law of iterated expectations, we have 
\begin{align}
        & \mathop{\mathbb{E}}[f((K+1)T)-f(KT)|\mathbf{X}(KT),\mathbf{Y}(KT),\mathbf{U}(KT)] \nonumber  \\
        & \leq M_{1} + M_{2} - \nonumber \\ 
        & 2\sum_{n = KT}^{\substack{(K+1)T \\ - 1}}\mathop{\mathbb{E}}[\mathop{\mathbb{E}}[W(\mathbf{X}(n),\mathbf{Y}(n),\mathbf{U}(n),s_{IID}(n),\bl^{\Delta_{IID}^{o}}(n))\nonumber \\
        &\mid \mathbf{X}(n),\mathbf{Y}(n),\mathbf{U}(n),\mathbf{X}(KT),\mathbf{Y}(KT),\mathbf{U}(KT)]]\nonumber \\
        & + 2T\sum_{i \in [N]}\lambda_{i}X_{i}(KT) - 2T\sum_{i \in [N]}\overline{P}_{i}U_{i}(KT) \nonumber \\
        & \leq M_{1} + M_{2} - \nonumber \\ 
        & 2\sum_{n = KT}^{\substack{(K+1)T \\ - 1}}\mathop{\mathbb{E}}[\mathop{\mathbb{E}}[W(\mathbf{X}(n),\mathbf{Y}(n),\mathbf{U}(n),s_{IID}(n),\bl^{\Delta_{IID}}(n))\nonumber \\
        &\mid \mathbf{X}(n),\mathbf{Y}(n),\mathbf{U}(n),\mathbf{X}(KT),\mathbf{Y}(KT),\mathbf{U}(KT)]]\nonumber \\
        \label{eq:inequality2}
        & + 2T\sum_{i \in [N]}\lambda_{i}X_{i}(KT) - 2T\sum_{i \in [N]}\overline{P}_{i}U_{i}(KT).
\end{align}
Note that (\ref{eq:inequality2}) was obtained using (\ref{eq:inequality1}). For every $\delta > 0$ in (\ref{eq:DeltaEqn1})-(\ref{eq:DeltaEqn3}), we have
\begin{align}
    & \mathop{\mathbb{E}}[f((K+1)T)-f(KT)|\mathbf{X}(KT),\mathbf{Y}(KT),\mathbf{U}(KT)] \nonumber  \\   
    & \leq M_{1} + M_{2} - \nonumber \\ 
    & 2\sum_{n = KT}^{\substack{(K+1)T \\ - 1}}\mathop{\mathbb{E}}\Bigg[\sum_{i \in [N]}[ X_{i}(n)(\lambda_{i} + \delta) + Y_{i}(n)\delta - U_{i}(n)(\oP_{i} - \delta) ]\nonumber \\
    &\mid \mathbf{X}(n),\mathbf{Y}(n),\mathbf{U}(n),\mathbf{X}(KT),\mathbf{Y}(KT),\mathbf{U}(KT)]\Bigg]\nonumber \\
    & + 2T\sum_{i \in [N]}\lambda_{i}X_{i}(KT) - 2T\sum_{i \in [N]}\overline{P}_{i}U_{i}(KT). \nonumber 
\end{align}
Let $\frac{M_{3}}{2} \myeq T^{2}\mu_{max}\sum_{}i\in [N](\lambda_{i}+\delta) + T^{2}\sum_{i\in [N]}\oP_{i}(\oP_{i}-\delta)$. Using (\ref{eq:Xrel2}) and (\ref{eq:Urel1}), we get
\begin{align}
    & \mathop{\mathbb{E}}[f((K+1)T)-f(KT)|\mathbf{X}(KT),\mathbf{Y}(KT),\mathbf{U}(KT)] \nonumber  \\ 
    \label{eq:inequality5}
    & \leq M_{1} + M_{2} + M_{3} -  2\delta\sum_{i \in [N]}(X_{i}(KT) + Y_{i}(KT) + U_{i}(KT)).
\end{align}
Let $\mathcal{B} = \{\{\mathbf{X}(KT),\mathbf{Y}(KT),\mathbf{U}(KT)\}:\sum_{i \in [N]}( X_{i}(KT) + Y_{i}(KT) +  U_{i}(KT) ) \leq (M_{1}+ M_{2}+M_{3} + 1)/2\delta \}$. Then,
\begin{align}
    \mathop{\mathbb{E}}& [f((K+1)T)-f(KT)|\mathbf{X}(KT),\mathbf{Y}(KT),\mathbf{U}(KT)] \nonumber \\
    &< \begin{cases}
        \infty, & \forall \text{ } \{\mathbf{X}(KT),\mathbf{Y}(KT),\mathbf{U}(KT)\} \\
        -1, & \text{if } \{\mathbf{X}(KT),\mathbf{Y}(KT),\mathbf{U}(KT)\} \notin \mathcal{B}. 
  \end{cases}
\end{align}
Thus, since $\mathopen|\mathcal{B}\mathclose|$ is finite, by Foster's Theorem \cite[Theorem~2.2.3]{GVM}, $\{\mathbf{X}(KT), \mathbf{Y}(KT), \mathbf{U}(KT)\}_{KT \geq 1}$ is positive recurrent, and for each queue the expected queue length under its stationary distribution is finite. Thus, the system is stable under $\Delta_{IID}^{T}$.

\section{Proof of Theorem 2}
\label{sec:ProofRWoptimal}
Let $\bm{\lambda}$ be stabilizable and 
\begin{align}
\label{eq:Lyamod2}
    f(n) = \sum_{i=1 \in [N]}\bigg[(X_{i}(n))^{2} + (Y_{i}(n))^{2} + (U_{i}(n))^{2}\bigg].
\end{align}
Let $M_{4}\myeq N[(TA_{max})^2 + (T\mu_{max})^2 + (N^2 + 1)T^2\mu_{max}^2 + 2T^2 \overline{P}_{i}^2]$. Using an analysis similar to that used for obtaining (\ref{eq:inequality3})
\begin{align}
        & \mathop{\mathbb{E}}[f((K+1)T)-f(KT)|\mathbf{X}(KT),\mathbf{Y}(KT),\mathbf{U}(KT)] \nonumber  \\
        & \leq   M_{4} - 2\times \nonumber \\ 
        & \sum_{n = KT}^{\substack{(K+1)T \\- 1}}\mathop{\mathbb{E}}\bigg[W(\mathbf{X}(KT),\mathbf{Y}(KT),\mathbf{U}(KT),s_{RW}(n),\bl^{\Delta_{RW}^{T}}(n)) \nonumber \\
        &\hspace{4cm}\mid \mathbf{X}(KT),\mathbf{Y}(KT),\mathbf{U}(KT)\Bigg] \nonumber \\
        \label{eq:inequality4}
        & + 2T\sum_{i \in [N]}\lambda_{i}X_{i}(KT) - 2T\sum_{i \in [N]}\overline{P}_{i}U_{i}(KT),
\end{align}
where, 
\begin{align}
    &W(\mathbf{X}(KT),\mathbf{Y}(KT),\mathbf{U}(KT),s_{RW}(n),\bl^{\Delta_{RW}^{T}}(n))  = \nonumber \\
    & \sum_{i \in [N]}\Bigg[\sum_{\substack{j \in [N]\\j\neq i}}\bigg( X_{i}(KT) - Y_{i}(KT) \bigg)\mu_{ij}(s_{RW}(n),\bl^{\Delta_{RW}^{T}}(n)) \nonumber \\
    &  + \max(X_{i}(KT),Y_{i}(KT))\mu_{i0}(s_{RW}(n),\bl^{\Delta_{RW}^{T}}(n)) \nonumber \\
    &  - U_{i}(KT)\sum_{j=0,j\neq i}^{N}\l_{ij}^{\Delta_{RW}^{T}}(n)\Bigg]. \nonumber 
\end{align}

Consider two parallel networks with one network having topology evolving in IID fashion and the other network having topology evolving as generalized random walk. Assume that the stationary distribution of the network with topology evolving as generalized random walk is same as IID distribution of the first network. Let $s_{IID}(n)$ and $s_{RW}(n)$ be the state of the grid under IID and Markov mobility process in slot $n$, respectively. If in a slot $n$, $s_{IID}(n) = s_{RW}(n)$, then
\begin{align}
\label{eq:Wcomp2}
    &W(\mathbf{X}(KT),\mathbf{Y}(KT),\mathbf{U}(KT),s_{IID}(n),\bl^{\Delta_{IID}^{T}}(n)) \nonumber \\
    &= W(\mathbf{X}(KT),\mathbf{Y}(KT),\mathbf{U}(KT),s_{RW}(n),\bl^{\Delta_{RW}^{T}}(n)).
\end{align}
Suppose in slot $n$, $s_{IID}(n) \neq s_{RW}(n)$. Towards that end, consider $\mu_{max}$ be the maximum transmission rate achievable across any pair of transmitter and receiver under any network topology. Also, let $P_{max}$ be the maximum power allowed at any MS in the network. Then, considering the assumption that each MS can transmit to only one receiver in maximum, we have the following relations:
\begin{align}
        \label{eq:MarkovRel1}
        & \mu_{ij}(s_{IID}(n),\bl^{\Delta_{IID}^{T}}(n)) \leq \mu_{ij}(s_{RW}(n),\bl^{\Delta_{RW}^{T}}(n)) + \mu_{max}, \\
        \label{eq:MarkovRel2}        
        & \mu_{ij}(s_{IID}(n),\bl^{\Delta_{IID}^{T}}(n)) + \mu_{max} \geq \mu_{ij}(s_{RW}(n),\bl^{\Delta_{RW}^{T}}(n)),  \\
        \label{eq:MarkovRel3}        
        & \mu_{i0}(s_{IID}(n),\bl^{\Delta_{IID}^{T}}(n)) \leq \mu_{i0}(s_{RW}(n),\bl^{\Delta_{RW}^{T}}(n)) + \mu_{max}, \\
        \label{eq:MarkovRel4}        
        & \max{(X_{i}(KT),Y_{i}(KT))} \leq X_{i}(KT) + Y_{i}(KT),  \\
        \label{eq:MarkovRel5}        
        & \sum_{j=0,j\neq i}^{N}\l_{ij}^{\Delta_{RW}^{T}}(n) + P_{max} \geq \sum_{j = 0, j\neq i}^{N}\l_{ij}^{\Delta_{IID}^{T}}(n), \\
        \label{eq:MarkovRel6}        
        & \sum_{j=0,j\neq i}^{N}\l_{ij}^{\Delta_{RW}^{T}}(n)  \leq \sum_{j = 0, j\neq i}^{N}\l_{ij}^{\Delta_{IID}^{T}}(n) + P_{max}.
\end{align}
Thus, for slots $n$ with $s_{IID}(n) \neq s_{RW}(n)$, using (\ref{eq:MarkovRel1})-(\ref{eq:MarkovRel6}), we have
\begin{align}
       &W(\mathbf{X}(KT),\mathbf{Y}(KT),\mathbf{U}(KT),s_{IID}(n),\bl^{\Delta_{IID}^{T}}(n)) \nonumber \\ 
       &\leq W(\mathbf{X}(KT),\mathbf{Y}(KT),\mathbf{U}(KT),s_{RW}(n),\bl^{\Delta_{RW}^{T}}(n)) \nonumber \\
       &+ \mu_{max}(N+1)\bigg(\sum_{i \in [N]}X_{i}(KT) + \sum_{i \in [N]}Y_{i}(KT)\Bigg) \nonumber \\
       &+ P_{max}\sum_{i \in N}U_{i}(KT).
\end{align}
For tractability, suppose there are two possible states, $s_1$ and $s_2$, that network topology can assume. Let number of mismatches in number of states when network evolves in IID fashion and when network topology evolves as a generalized random walk in a single time interval of $T$ slots be $\tau$. That is,
\begin{equation}
\begin{split}
    \tau = & \mid \text{number of states } s_1 \text{ in IID process } \\
    &- \text{number of states } s_1 \text{ in Markov process}\mid. \nonumber
\end{split}
\end{equation}
Let $\gamma = (N+1)\mu_{max}\sum_{i \in [N]}X_{i}(KT) + (N+1)\mu_{max}\sum_{i \in [N]}Y_{i}(KT) + P_{max}\sum_{i \in N}U_{i}(KT)$. Thus, in (\ref{eq:inequality4}), we have
\begin{align}
        & \mathop{\mathbb{E}}[f((K+1)T)-f(KT)|\mathbf{X}(KT),\mathbf{Y}(KT),\mathbf{U}(KT)]   \nonumber \\
        & \leq M_{4} - 2\times\nonumber \\ 
        & \sum_{n = KT}^{\substack{(K+1)T \\ -1}}\mathop{\mathbb{E}}[W(\mathbf{X}(KT),\mathbf{Y}(KT),\mathbf{U}(KT),s_{IID}(n),\bl^{\Delta_{IID}^{T}}(n))\nonumber \\
        &\hspace{4cm}|\mathbf{X}(KT),\mathbf{Y}(KT),\mathbf{U}(KT)]  \nonumber \\
        &+ \gamma\mathop{\mathbb{E}}[\tau]+ 2T\sum_{i \in [N]}\lambda_{i}X_{i}(KT) -2T\sum_{i \in [N]}\overline{P}_{i}U_{i}(KT).
\end{align}
Now for some $\epsilon>0$ and $\delta>0$, we can write 
\begin{align}
\label{eq:DeltaEqn4}
& \sum_{s\in\cS}\sum_{k=1}^M\pi_{s}w_{sk} \left[ \mu_{i0}(s,\l_{k})q_{isk}+\sum_{j\in [N], j \neq i}\mu_{ij}(s,\l_{k})\right] \nonumber \\
& \hspace{3cm}\geq  \lambda_{i} + \delta + \epsilon, \\
\label{eq:DeltaEqn5}
&\sum_{s\in\cS}\sum_{k=1}^M\pi_{s}(1-q_{isk})w_{sk}\mu_{i0}(s,\l_{k}) \quad \nonumber \\
& \hspace{2cm}\geq \sum_{s\in\cS}\sum_{k=1}^M\sum_{j\in [N], j \neq i}\pi_{s}w_{sk}\mu_{ji}(s,\l_{k}) + \delta + \epsilon,\\
\label{eq:DeltaEqn6}
&\sum_{s\in\cS}\sum_{k=1}^M\pi_{s}w_{sk}\sum_{j\in[N], j \neq i}\l_{ijk} + \delta + \epsilon \leq \oP_{i}.  
\end{align}
Using (\ref{eq:DeltaEqn4})-(\ref{eq:DeltaEqn6}) and with similar steps as taken in obtaining (\ref{eq:inequality5}) from (\ref{eq:inequality3}), we can show that
\begin{align}
    & \mathop{\mathbb{E}}[f((K+1)T)-f(KT)|\mathbf{X}(KT),\mathbf{Y}(KT),\mathbf{U}(KT)] \nonumber  \\ 
    & \leq M_{4} + M_{2} + M_{3} \nonumber \\ 
    & + \big((N+1)\mu_{max}\mathop{\mathbb{E}}[\tau]-2\epsilon\big)\sum_{i\in [N]}X_{i}(KT) \nonumber \\
    & + \big((N+1)\mu_{max}\mathop{\mathbb{E}}[\tau]-2\epsilon\big)\sum_{i\in [N]}Y_{i}(KT) \nonumber \\
    & + \big(P_{max}\mathop{\mathbb{E}}[\tau]-2\epsilon\big)\sum_{i\in [N]}U_{i}(KT) \nonumber \\
    & -  2\delta\sum_{i \in [N]}(X_{i}(KT) + Y_{i}(KT) + U_{i}(KT)).
\end{align}
Note that if $\epsilon > \frac{1}{2}\max\{(N+1)\mu_{max}\mathop{\mathbb{E}}[\tau],P_{max}\mathop{\mathbb{E}}[\tau]\}$, then, we have 
\begin{align}
    & \mathop{\mathbb{E}}[f((K+1)T)-f(KT)|\mathbf{X}(KT),\mathbf{Y}(KT),\mathbf{U}(KT)] \nonumber  \\ 
    \label{eq:inequality6}
    & \leq M_{4} + M_{2} + M_{3} \nonumber \\ 
    & -  2\delta\sum_{i \in [N]}(X_{i}(KT) + Y_{i}(KT) + U_{i}(KT)).    
\end{align}
Using exponentially fast convergence of empirical distribution
to the unique stationary distribution for ergodic Markov chains \cite{AA}, we have $\mathop{\mathbb{E}}[\tau] \rightarrow 0$ as $T \rightarrow \infty$. Let $\mathcal{C_{\epsilon}} = \{\{\mathbf{X}(KT),\mathbf{Y}(KT),\mathbf{U}(KT)\}:\sum_{i \in [N]}( X_{i}(KT) + Y_{i}(KT) +  U_{i}(KT) ) \leq (M_{1}+ M_{2}+M_{3} +  1)/2\delta \}$. Then,
\begin{align}
    \mathop{\mathbb{E}}& [f((K+1)T)-f(KT)|\mathbf{X}(KT),\mathbf{Y}(KT),\mathbf{U}(KT)] \nonumber \\
    &< \begin{cases}
        \infty, & \forall \text{ } \{\mathbf{X}(n),\mathbf{Y}(n),\mathbf{U}(n)\} \\
        -1, & \text{if } \{\mathbf{X}(n),\mathbf{Y}(n),\mathbf{U}(n)\} \notin \mathcal{C}_{\epsilon}. 
  \end{cases}
\end{align}
For any $\epsilon >  \frac{1}{2}\max\{(N+1)\mu_{max}\mathop{\mathbb{E}}[\tau],P_{max}\mathop{\mathbb{E}}[\tau]\}$, we have finite $\mathopen|\mathcal{C}_{\epsilon}\mathclose|$. By Foster's Theorem \cite[Theorem~2.2.3]{GVM}, $\{\mathbf{X}(KT), \mathbf{Y}(KT), \mathbf{U}(KT)\}_{KT \geq 1}$ is positive recurrent, and for each queue the expected queue length under its stationary distribution is finite. Summarizing, depending on size of the interval $T$, we decide on some $\epsilon>\frac{1}{2}\max\{(N+1)\mu_{max}\mathop{\mathbb{E}}[\tau],P_{max}\mathop{\mathbb{E}}[\tau]\}$ which guarantees that the system is stable under policy $\Delta_{RW}^{T}$ $\forall \bm{\lambda} \in \Lambda_{\epsilon}$. 